# Optical Monitoring and Long-term Optical Spectral Variability of BL Lacertae Object S5 0716+714

Huai-Zhen Li[1], Di-Fu Guo[2], Long-Hua Qin[1], Fen Liu[2], Hong-Tao Liu[3,4,5], Ting-Feng Yi[6], Quan-Gui Gao[1], Shi-Feng Huang[7,8], Xing Gao[9,10], and Xu Chen[2]
[1] Physics Department, Yuxi Normal University, Yuxi 653100, Yunnan, People's Republic of China; qinlh@mail.ustc.edu.cn
[2] Shandong Provincial Key Laboratory of Optical Astronomy and Solar-Terrestrial Environment, Institute of Space Sciences, Shandong University, Weihai 264209, Shandong, People's Republic of China; difu@sdu.edu.cn, liufen210@sdu.edu.cn
[3] Yunnan Observatories, Chinese Academy of Sciences, 396 Yangfangwang, Guandu District, Kunming 650216, Yunnan, People's Republic of China; htliu@ynao.ac.cn
[4] Key Laboratory for the Structure and Evolution of Celestial Objects, Chinese Academy of Sciences, Kunming 650216, Yunnan, People's Republic of China
[5] Center for Astronomical Mega-Science, Chinese Academy of Sciences, 20A Datun Road, Chaoyang District, Beijing 100012, People's Republic of China
[6] Physics Department, Yunan Normal University, Kunming 650092, Yunnan, People's Republic of China
[7] CAS Key Laboratory for Research in Galaxies and Cosmology, Department of Astronomy, University of Science and Technology of China, Hefei 230026, Anhui, People's Republic of China
[8] School of Astronomy and Space Sciences, University of Science and Technology of China, Hefei 230026, Anhui, People's Republic of China
[9] Xinjiang Astronomical Observatory, 150 Science 1-Street, Urumqi 830011, Xinjiang, People's Republic of China
[10] Urumqi No.1 Senior High School, Urumqi 830002, Xinjiang, People's Republic of China
Received 2024 July 22; revised 2025 January 11; accepted 2025 January 24; published 2025 February 25

## Abstract

We present photometric observations of the BL Lacertae object S5 0716+714 with a temporal resolution of 120 s in the Sloan $i'$ and $r'$ bands. These observations were conducted using the Comet Search Program telescope at Xingming Observatory from 2018 December 22 to 2020 February 15, and more than 5600 effective images were obtained on each filter across 79 nights. Additionally, we compiled long-term variability data spanning 34 yr in the optical UBVRI bands. Using the power-enhanced F-test and nested ANOVA test, we found intraday variability (IDV) on 31 nights and possible IDV on 20 nights in the $i'$ band. Similarly, IDV was detected on 35 nights in the $r'$ band, while possible IDV was observed on 22 nights. The minimum variability timescale is $7.33 \pm 0.74$ minutes, and the estimated black hole masses are $(0.68 \sim 5.12) \times 10^8 M_\odot$. The spectral variability and long-term optical light curves reveal a bluer-when-brighter trend on intraday timescales. The long-term optical flux density and spectral index exhibit periodic variability with a timescale of about 1038 days. An anticorrelation between optical flux and spectral index was observed, with a time delay of $-140$ days. Variability across different optical bands exhibited a strong correlation, with no discernible time lag. From the IDV, spectral variability, correlation, and time delays between different bands, we conclude that these radiation characteristics may result from the shock-in-jet model scenario.

*Unified Astronomy Thesaurus concepts:* Blazars (164)

*Materials only available in the online version of record:* machine-readable tables

## 1. Introduction

Blazars are a special subclass of active galactic nuclei with their relativistic jets oriented almost along the line of sight of the observer, leading to the Doppler-boosted emission from the jet (R. D. Blandford & A. Königl 1979; C. M. Urry & P. Padovani 1995; N. Kaur et al. 2018). They radiate across the entire electromagnetic spectrum from radio radiation to γ-rays with rapid and violent variability at almost all wave bands. Therefore, the variability analysis is a tool to probe the emission properties of blazars (C. M. Urry & P. Padovani 1995). Blazars consist of two subclasses: flat spectrum radio quasars (FSRQs) and BL Lacertae objects are distinguished by their emission lines. FSRQs have broad emission lines with an equivalent width (EW) greater than 5 Å, while BL Lacertae objects have weak or no broad emission lines, with an EW less than 5 Å (J. T. Stocke et al. 1991; M. J. M. Marcha et al. 1996). In addition, the spectral energy distribution (SED) of blazars shows two prominent peaks due to nonthermal radiation (G. Fossati et al. 1998). The low-energy peak spans from the infrared to X-rays, and the high-energy peak lies within the MeV to TeV γ-ray range (G. Fossati et al. 1998). The low-energy peak results from synchrotron radiation, while the high-energy peak arises from inverse Compton scattering, hadronic processes, or both (e.g., A. P. Marscher & W. K. Gear 1985; G. Ghisellini et al. 1998, 2010; C. D. Dermer et al. 2012; M. Böttcher et al. 2013; L. Chen 2018). Depending on the SED, BL Lacertae objects are commonly classified into two or three subclasses (A. A. Abdo et al. 2010). This classification scheme was first proposed by P. Padovani & P. Giommi (1995), who used the peak energy of synchrotron emission to categorize BL Lacertae objects into low-energy synchrotron peak BL Lacertae objects and high-energy synchrotron peak BL Lacertae objects. This classification was later extended to blazars by A. A. Abdo et al. (2010). Blazars are further classified based on their synchrotron peak frequencies into low-synchrotron-peaked blazars ($\nu^p_{\mathrm{syn}} < 10^{14}$ Hz), intermediate-synchrotron-peaked blazars ($10^{14}$ Hz $< \nu^p_{\mathrm{syn}} < 10^{15}$ Hz), and high-energy-peaked blazars ($\nu^p_{\mathrm{syn}} > 10^{15}$ Hz) (A. A. Abdo et al. 2010).

Blazars exhibit variability over timescales from minutes to decades (e.g., N. Kaur et al. 2018; H.-C. Feng et al. 2020a; H.-Z. Li et al. 2024a). This variability is categorized into intraday variability (IDV) or microvariability (minutes to hours), short-term variability (STV) (days to months), and long-term variability (LTV) (months to years) (e.g., S. J. Wagner & A. Witzel 1995; J.-H. Fan et al. 2009; H. Z. Li et al. 2016; N. Kaur et al. 2018; H.-Z. Li et al. 2024a). The mechanism behind IDV, in particular, remains debated. LTV may be linked to the orbital motion of the black hole, jet structure changes, or accretion disk instabilities (A. P. Marscher & W. K. Gear 1985; T. Kawaguchi et al. 1998; G. Z. Xie et al. 2002, 2008; H. Yang et al. 2023). STV involves intrinsic and extrinsic factors such as shocks in the jet, fresh plasma injection, variations in the Doppler factor, and gravitational microlensing (e.g., N. Kaur et al. 2018; D. Xiong et al. 2020; H.-Z. Li et al. 2022b; H. Yang et al. 2023). IDV could result from plasma compression by shocks, shock interactions with inhomogeneities, Kelvin–Helmholtz instability, and variability in the boosting factor (e.g., G. E. Romero 1995; G. Ghisellini et al. 1997; M. Villata et al. 2002; H. Poon et al. 2009; S. Hong et al. 2017; D. Xiong et al. 2020; J. Fan et al. 2023). Furthermore, IDV near the supermassive black hole allows estimation of the radiation region size and black hole mass based on the minimum variability timescale (H. R. Miller et al. 1989; H. T. Liu & J. M. Bai 2015; D. Xiong et al. 2017; X. Chang et al. 2023).

Blazar emissions are typically accompanied by concurrent spectral or color variations (e.g., H. Poon et al. 2009; D. Xiong et al. 2020; M. A. Gorbachev et al. 2022; X. Chang et al. 2023). Examining these color changes with flux variability helps deduce the underlying mechanisms (M. A. Gorbachev et al. 2022). Color variations in blazars include bluer-when-brighter (BWB), redder-when-brighter (RWB), or achromatic trends (e.g., H. Poon et al. 2009; S. Hong et al. 2017; D. Xiong et al. 2020; M. A. Gorbachev et al. 2022; X. Chang et al. 2023; H.-Z. Li et al. 2024a). Most blazars exhibit the BWB trend, while the RWB trend is often seen in low-flux-state FSRQs (D. Xiong et al. 2020). The emergence of the BWB trend primarily arises from nonthermal emissions originating from the relativistic jet, while the manifestation of BWB behavior can also be elucidated within the framework of the accretion disk model when the thermal component emanating from the accretion disk dominates the overall radiation (M. F. Gu & S.-L. Li 2013; H. Liu et al. 2016; D. Xiong et al. 2020). Additionally, the chromatism of BWB can be elucidated by the contribution of two components with "red" and "blue" colors to the observed radiation flux (M. A. Gorbachev et al. 2022). The BWB trend can be observed as the relative contributions of these red and blue components to the total flux undergo changes. The red component is associated with the synchrotron radiation emitted by the jet, while the blue component is linked to the thermal radiation originating from the accretion disk (M. F. Gu et al. 2006; J. C. Isler et al. 2017; M. A. Gorbachev et al. 2022). However, the two-component interpretation cannot be applied to explain the color behavior of BL Lacertae objects due to their quasi-featureless spectrum (M. A. Gorbachev et al. 2022). The RWB chromatic, commonly observed in low-flux-state FSRQs, arises from a combination of accretion disk and jet components (M. F. Gu et al. 2006; D. Xiong et al. 2020). Moreover, achromatic color trends can arise from variations in the Doppler factor (M. Villata et al. 2004; S. M. Hu et al. 2014; D. Xiong et al. 2020).

S5 0716+714, classified as an intermediate-synchrotron-peaked blazar (A. A. Abdo et al. 2010), is one of the most active blazars across all energy bands, with a redshift of $z = 0.31 \pm 0.08$. It is included in the catalog of TeV-emitting sources, with very high energy (VHE) γ-ray emissions detected by Major Atmospheric Gamma Imaging Cherenkov (MAGIC) observations (H. Anderhub et al. 2009). S5 0716+714 is among the brightest BL Lacertae objects, consistently showing high activity and considerable brightness, making it observable with moderate facilities (N. Kaur et al. 2018; MAGIC Collaboration et al. 2018). It is extensively studied for variability across the electromagnetic spectrum on diverse timescales (e.g., S. Wagner et al. 1990; A. Quirrenbach et al. 1991; S. J. Wagner et al. 1996; G. Ghisellini et al. 1997; C. M. Raiteri et al. 2003; G. Tagliaferri et al. 2003; L. Foschini et al. 2006; M. Villata et al. 2008; H. Poon et al. 2009; A. C. Gupta et al. 2009, 2012; G. Bhatta et al. 2013; S. Chandra et al. 2015; A. Wierzcholska & H. Siejkowski 2016; S. Hong et al. 2017; H. Z. Li et al. 2018; MAGIC Collaboration et al. 2018; H. T. Liu et al. 2019; D. Xiong et al. 2020; H. Yang et al. 2023; T. Tripathi et al. 2024). Optical IDV has been observed by many authors (e.g., M. Villata et al. 2000; J. Wu et al. 2005; F. Montagni et al. 2006; H. Poon et al. 2009; B.-z. Dai et al. 2015; A. Agarwal et al. 2016; S. Hong et al. 2017; H. T. Liu et al. 2019; E. Shablovinskaya & V. Afanasiev 2020; D. Xiong et al. 2020; C. M. Raiteri et al. 2021; T. Tripathi et al. 2024), with periodic or quasiperiodic oscillations detected on several occasions (A. C. Gupta et al. 2009; B. Rani et al. 2010; S. Hong et al. 2018). STV and LTV have also been extensively documented (e.g., A. C. Gupta et al. 2008; N. H. Liao et al. 2014; B.-z. Dai et al. 2015; Y.-H. Yuan et al. 2017; H. Z. Li et al. 2018; Y. Dai et al. 2021; X.-P. Li et al. 2023; H. Yang et al. 2023). The spectral variability of S5 0716+714 has been widely studied, with a strong BWB trend consistently reported (e.g., J. Wu et al. 2007; H. Poon et al. 2009; B.-z. Dai et al. 2015; N. Kaur et al. 2018; D. Xiong et al. 2020; T. Tripathi et al. 2024), though some studies suggest a mild or absent BWB trend (e.g., G. Ghisellini et al. 1997; C. M. Raiteri et al. 2003; S. M. Hu et al. 2014; A. Agarwal et al. 2016; S. Hong et al. 2017; X. Zhang et al. 2018). Estimates of the supermassive black hole (SMBH) mass at the center of S5 0716+714 range from $\sim 2.5 \times 10^6 M_\odot$ to $4.4 \times 10^{10} M_\odot$ (e.g., E. W. Liang & H. T. Liu 2003; X. Zhang et al. 2008; A. C. Gupta et al. 2009; B. K. Zhang et al. 2012; B.-z. Dai et al. 2015; A. Agarwal et al. 2016; G. Bhatta et al. 2016b; Y.-H. Yuan et al. 2017; S. Hong et al. 2018; N. Kaur et al. 2018; H. T. Liu et al. 2019; M. S. Butuzova 2021; X.-L. Liu et al. 2021). However, the SMBH mass could not be precisely determined due to the lack of optical measurements for the apparent velocity of the source ($\beta_{\text{app}}$) (D. Ł. Król et al. 2023).

This paper presents the results of a study on the emission properties of S5 0716+714 by analyzing the multiband optical flux and spectral variability. The emission from the source was observed using the Comet Search Program (CSP) telescope at Xingming Observatory in the $i'$ and $r'$ bands, while compiling long-term multiband optical variability data. The data was analyzed to study the source's IDV, spectral variability, variability timescale, and the correlation between the different bands. In Section 2, the observations and data reductions are presented. The analysis techniques are presented in Section 3. Results and discussion are shown in Section 4. Finally, the conclusions are given in Section 5.

## 2. Observations and Data Reductions

### 2.1. Optical Observations and Light Curves

From 2018 December 22 to 2020 February 15, BL Lacertae object S5 0716+714 was observed on 79 nights using the CSP telescope at the Xingming Observatory.[11] The Sloan $r'$ and $i'$

---
[11] http://xjltp.china-vo.org/

**Table 1**
Reference Stars

| Comparison Star | $r'_{\mathrm{mag}}$ [a] | $i'_{\mathrm{mag}}$ [a] |
|---|---|---|
| 1 | 10.890 | 10.798 |
| 5 | 13.367 | 13.229 |
| 8 | 13.991 | 13.899 |

**Note.**
[a] Magnitudes were taken from the catalog of APASS DR10 (A. A. Henden 2019).

bands were used during the observations. The exposure time for both filters is 120 s. More than 5600 effective images on each filter were obtained. The aperture of the CSP telescope is 10 cm, equipped with an Apogee U16M CCD, possessing 4096 × 4096 square pixels, offering a field of view of nearly 4° × 4°. All data were processed by bias, dark, and flat-field correction. The task APPHOT of the IRAF software package was used to do the photometry. Comparison stars 1, 5, and 8 taken from M. Villata et al. (1998) were used for calibration. The magnitude of the comparison stars in the Sloan $r'$ and $i'$ bands are displayed in Table 1. Then, the magnitude of the target was derived from differential photometry. The error is given as below,

$$\sigma = \sqrt{\frac{(m_1 - \overline{m})^2 + (m_5 - \overline{m})^2 + (m_8 - \overline{m})^2}{2}}, \quad (1)$$

where $m_1$, $m_5$, and $m_8$ is the magnitude of S5 0716+714 calibrated by the first, fifth, and eighth comparison star, respectively, whereas $\overline{m}$ is the averaged magnitude of S5 0716+714 obtained from the comparison stars. In the following section, we will analyze the observation to study the light curves, the IDV, the minimum variability timescale, the color behavior, and the correlation between the variability of the Sloan $i'$ and $r'$ bands.

The observation data and overall light curves for the Sloan $i'$ and $r'$ bands are given in Table 2 and Figure 1. The observation period exhibits a clear division into two distinct segments, as shown in Figure 1: from 2018 December 22 (MJD 8475) to 2019 February 25 (MJD 8540), spanning a duration of 65 days; and from 2019 December 8 (MJD 8826) to 2020 February 15 (MJD 8895), spanning approximately 69 days. The brightness exhibits a declining trend with an amplitude of approximately 2 mag during the initial observation period, whereas it initially increases and then decreases with an amplitude of about 1.3 mag during the subsequent observation period. The brightness of the second observation period surpasses that of the first. The light curves dividing into two segments were depicted in Figure 2, exhibiting discernible variability throughout the observation period.

### 2.2. Long-term Optical Variability Data

The variability data of S5 0716+714 in the optical UBVRI bands from 1991 to 2023 were compiled from the literature and databases. The data from the literature include those of G. Ghisellini et al. (1997), S. Katajainen et al. (2000), B. Qian et al. (2002), C. M. Raiteri et al. (2003), V. R. Amirkhanyan (2006), M. F. Gu et al. (2006), F. Montagni et al. (2006), J. Wu et al. (2007, 2012), A. C. Gupta et al. (2008), X. Zhang et al. (2008), H. Poon et al. (2009), S. Chandra et al. (2011), G. Bhatta et al. (2013), V. M. Larionov et al. (2013), F. Taris et al. (2013), N. H. Liao et al. (2014), B.-z. Dai et al. (2015), V. T. Doroshenko & N. N. Kiselev (2017), S. Hong et al. (2017, 2018), K. Nilsson et al. (2018), J. Xu et al. (2019), H. T. Liu et al. (2019), H.-C. Feng et al. (2020a), D. Xiong et al. (2020), M. A. Gorbachev et al. (2022), and L. Lu et al. (2024). The databases include the Multi-Optical-Band Polarization of Selected Blazars (MOBPOL) program,[12] the Steward Observatory blazar monitoring[13], and Swift's Ultraviolet/Optical Telescope (UVOT).[14] These variability data were observed using the Johnson–Cousins UBVRI system, the Sloan Digital Sky Survey (SDSS) system, and the Swift UVOT system. Based on the calibration equation presented by T. S. Chonis & C. M. Gaskell (2008) and W. Li et al. (2006), the data observed using the SDSS $r'i'$ filters and Swift UVOT *ubv* filters were calibrated to the standard Johnson–Cousins UBVRI system. The SDSS photometry for the $r'$ and $i'$ filters was calibrated by the Johnson–Cousins UBVRI system using the following equations (T. S. Chonis & C. M. Gaskell 2008):

$$R = r' - (0.272 \pm 0.092)(r' - i') - (0.159 \pm 0.22),$$
$$I = i' - (0.337 \pm 0.19)(r' - i') - (0.370 \pm 0.041). \quad (2)$$

Moreover, the Swift UVOT photometry for *u*, *b*, and *v* filters was calibrated the Johnson–Cousins UBVRI system using the following equations (W. Li et al. 2006):

$$U = V + 0.087 + 0.8926(u - v) + 0.0274(u - v)^2,$$
$$B = b + 0.0173 + 0.0187(u - b)$$
$$\quad + 0.013(u - b)^2 - 0.0108(u - b)^3$$
$$\quad - 0.0058(u - b)^4 + 0.0026(u - b)^5,$$
$$V = v + 0.0006 - 0.0113(b - v)$$
$$\quad + 0.0097(b - v)^2 - 0.0036(b - v)^3. \quad (3)$$

The UBVRI magnitudes of S5 0716+714 are corrected for foreground Galactic interstellar reddening using the color excess $E(B - V)$. For S5 0716+714, the color excess is $E(B - V)_{\mathrm{SFD}} = 0.031$ (D. J. Schlegel et al. 1998). Based on the value of $R_V$ for UBVRI bands given by E. F. Schlafly & D. P. Finkbeiner (2011) under the extinction law 3.1 of D. J. Schlegel et al. (1998), the Galactic extinction coefficients were estimated as $\Delta m_U = 0.134$ mag, $\Delta m_B = 0.112$ mag, $\Delta m_V = 0.085$ mag, $\Delta m_R = 0.067$ mag, and $\Delta m_I = 0.047$ mag and the UBVRI magnitudes were corrected for Galactic extinction. To study the long-term behavior of the optical spectral index variability, the magnitude is converted to optical flux density using the equation $F = F_0 10^{-0.4m}$, where $F_0$ represents the zero-magnitude equivalent flux density. The paper utilizes the value of $F_0$ provided by A. R. G. Mead et al. (1990). Additionally, the daily-averaged flux density is calculated to mitigate the heavy weighting of observational data and to reduce small amplitude intrahour fluctuations (Y. G. Zheng et al. 2008). The daily-averaged light curves of the UBVRI bands are presented in Figure 3, indicating that S5 0716+714 exhibits very actively in the optical band.

---

[12] https://www.bu.edu/blazars/mobpol/mobpol.html
[13] http://james.as.arizona.edu/~psmith/Fermi/DATA/Rphotdata.html
[14] https://swift.gsfc.nasa.gov/about_swift/uvot_desc.html

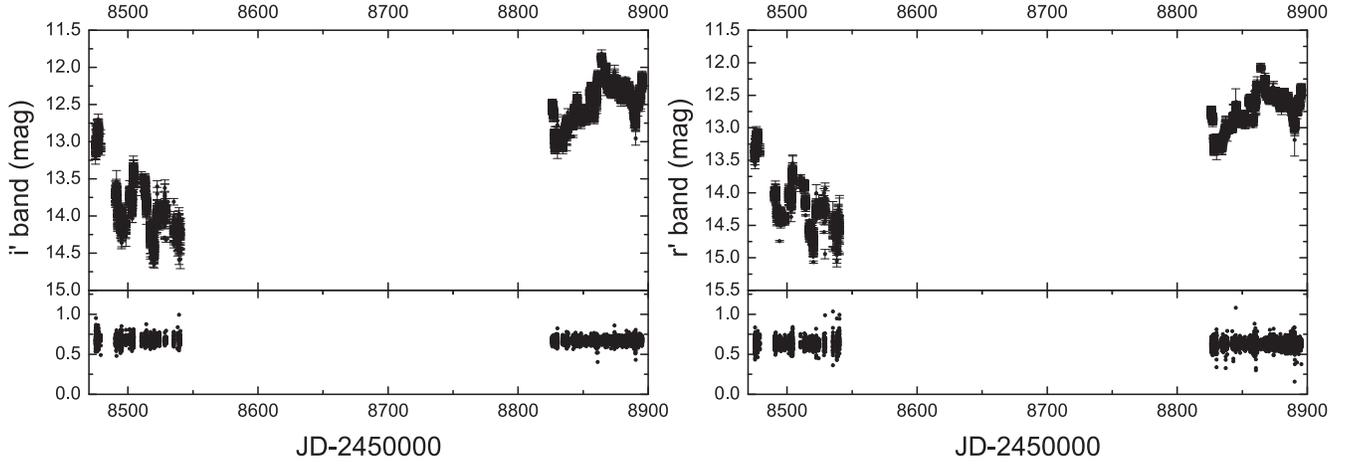

**Figure 1.** The light curves of S5 0716+714 in the Sloan $i'$ and $r'$ bands. The difference in magnitude between the two comparison stars 5 and 8 are displayed at the bottom of the corresponding panel.

**Table 2**
The Log of Observations of S5 0716+714 in the Sloan $i'$ and $r'$ Bands

| | $i'$ Band | | | | $r'$ Band | | |
|---|---|---|---|---|---|---|---|
| Date | MJD | Mag | $\sigma$ | Date | MJD | Mag | $\sigma$ |
| (1) | (2) | (3) | (4) | (5) | (6) | (7) | (8) |
| 2018-12-22 | 8475.00862 | 13.144 | 0.085 | 2018-12-22 | 8475.00470 | 13.531 | 0.04 |
| 2018-12-22 | 8475.01255 | 13.087 | 0.011 | 2018-12-22 | 8475.01060 | 13.338 | 0.027 |
| 2018-12-22 | 8475.01648 | 13.116 | 0.041 | 2018-12-22 | 8475.01454 | 13.367 | 0.072 |

**Note.** The initial three observed data points are presented herein, while the comprehensive table is available in a machine-readable format.

(This table is available in its entirety in machine-readable form in the online article.)

Furthermore, the variability among the UBRVI bands remains consistent.

### 3. Analysis Techniques

To investigate the variability properties of S5 0716+714, we employed the power-enhanced F-test and nested analysis of variance (ANOVA) test (J. A. de Diego 2014; J. A. de Diego et al. 2015; A. Pandey et al. 2019; T. Tripathi et al. 2024) to explore IDV. These methods have been developed to enhance the statistical power and reliability of the analysis, aiming at improving the detection of microvariability in blazar. The aforementioned techniques demonstrate superior reliability and robustness compared to commonly used statistical tests such as the C-test or F-test (T. Tripathi et al. 2024). Moreover, the percentage of amplitude change in the magnitude is also a crucial parameter for characterizing the variability.

#### 3.1. Power-enhanced F-test

The power-enhanced F-test is a statistical procedure utilized for detecting microvariability in blazar differential light curves by comparing the variance of the blazar's light curve with the combined variance of multiple comparison stars. This method enhances the power of the traditional F-test by incorporating multiple comparison stars, thereby mitigating non-normality effects in the data and increasing test reliability. In order to detect microvariability, it is imperative to carefully select a suitable comparative star as a reference and subsequently compare the variance of the blazar light curve with the combined variance exhibited by the comparison stars. The formula for the power-enhanced F-test is given as follows:

$$F_{\text{enh}} = \frac{s_{\text{blz}}^2}{s_c^2}, \quad (4)$$

where $s_{\text{blz}}^2$ represents the variance of differential light curves between instrumental magnitudes of blazar and reference star, and $s_c^2$ denotes the variance of combined differential LCs obtained from instrumental magnitudes of comparison star and reference star. The calculation for combined variance $s_c^2$ can be expressed as

$$s_c^2 = \frac{1}{(\sum_{j=1}^k N_j) - k} \sum_{j=1}^k \sum_{i=1}^{N_i} s_{j,i}^2, \quad (5)$$

where $N_j$ signifies the number of data points for $j$th comparison star, while $k$ denotes the count of comparison stars excluding the reference star. On the other hand, $s_{j,i}^2$ represents scaled square deviation for $j$th comparison star defined as

$$s_{j,i}^2 = \omega_j (m_{j,i} - \bar{m}_j)^2, \quad (6)$$

where the scaling factor $\omega_j$ is determined by the ratio of the average squared error of the blazar's differential light curve to that of the $j$th comparison star, $m_{j,i}$ represents a representation of the differential magnitude and $\bar{m}_j$ denotes a representation of the mean magnitude of the $j$th comparison star. The degrees of freedom in the numerator and denominator for the power-enhanced F-test are $\nu_1 = N - 1$, and $\nu_2 = k(N - 1)$, respectively, with $N$ representing the number of observations. It should be noted that blazar and all comparison stars have an equal number of observations. By utilizing the power-enhanced F-test, one can derive the critical value of the $F_{\mathrm{enh}}(99)$ that corresponds to a significance level of 0.01. The blazar is considered to exhibit variability with 99% confidence when the value of $F_{\mathrm{enh}}$ estimated from Equation (4) exceeds the critical value of $F_{c,\mathrm{enh}}$. The critical value of $F_{c,\mathrm{enh}}$ can be obtained as $F^{\alpha}_{\nu_1,\nu_2}$, where $\alpha$ represents the significance level ($\alpha = 0.01$). In this work, the comparison stars C1, C5, and C8 is used. The comparison star C5 was used as a reference star as it is closest in magnitude and color to the blazar. Hence, $k = 2$ since there are still two remaining comparison stars.

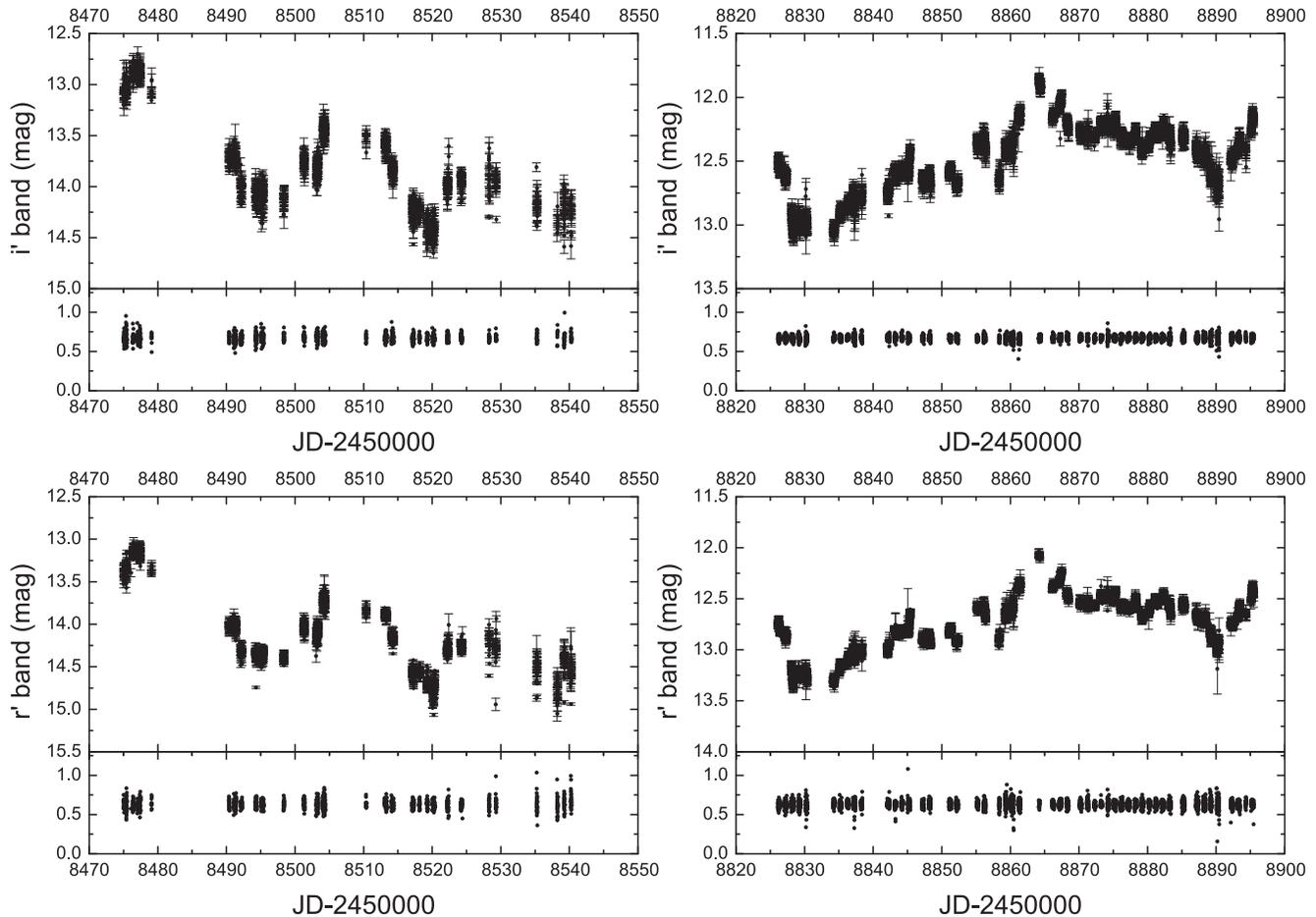

**Figure 2.** The light curve is partitioned into two segments based on the time of observation. The difference in magnitude between the two comparison stars 5 and 8 are displayed at the bottom of the corresponding panel.

### 3.2. Nested ANOVA Test

The nested ANOVA, an enhanced version of the traditional ANOVA, is employed to analyze blazar differential light curves by incorporating multiple stars. This approach proves particularly valuable when there is a limited availability of bright stars in the blazar field. The nested ANOVA investigates variances across three analysis stages: differences between groups (blazar variability), differences between observations due to shot noise and sky subtraction, and variance caused by different reference stars. The formula for the F-statistic in nested ANOVA can be expressed as follows:

$$F = \frac{\mathrm{MS_G}}{\mathrm{MS_{O(G)}}}, \qquad (7)$$

where $\mathrm{MS_G}$ represents the mean square due to groups and $\mathrm{MS_{O(G)}}$ denotes the mean square due to nested observations within groups. The degrees of freedom for each mean square are calculated based on the number of groups, observations, and stars utilized in the analysis. The mean squares $\mathrm{MS_G}$ and $\mathrm{MS_{O(G)}}$ are obtained by dividing the sums of squares $\mathrm{SS_G}$ (sum of squares due to groups) and $\mathrm{SS_{O(G)}}$ (sum of squares due to nested observations in groups), respectively, by their corresponding degrees of freedom $\nu$. The values for $\mathrm{SS_G}$ and $\mathrm{SS_{O(G)}}$ were estimated using Equation (4) from J. A. de Diego et al. (2015). The degrees of freedom between groups $\nu_1$ and nested observations in groups $\nu_2$ are $\nu_1 = a - 1$ and $\nu_2 = a(b - 1)$, respectively, where $a$ is the number of groups in the night's observations and $b$ is the number of data points in each group. We utilize comparison stars C1, C5, and C8 as reference sources to generate the differential light curves of the blazar. Subsequently, we divide these differential light curves into groups of five data points each, with the exception of the final group (J. A. de Diego 2010; J. A. de Diego et al. 2015; D. Xiong et al. 2016; S. Hong et al. 2017; T. Tripathi et al. 2024). The final group consists of $N - 5(a - 1)$ data points, ranging from 5 to 10, where $N$ represents the total data set size. Nested ANOVA involves a critical value $F_{c,\text{ANOVA}} = F^{\alpha}_{\nu_1,\nu_2}$, which is also present in the power-enhanced F-test. When $\alpha = 0.01$, if the F-value obtained from Equation (7) exceeds the critical value $F_{c,\text{ANOVA}}$, it indicates variability in the blazar with a confidence level higher than 99%.

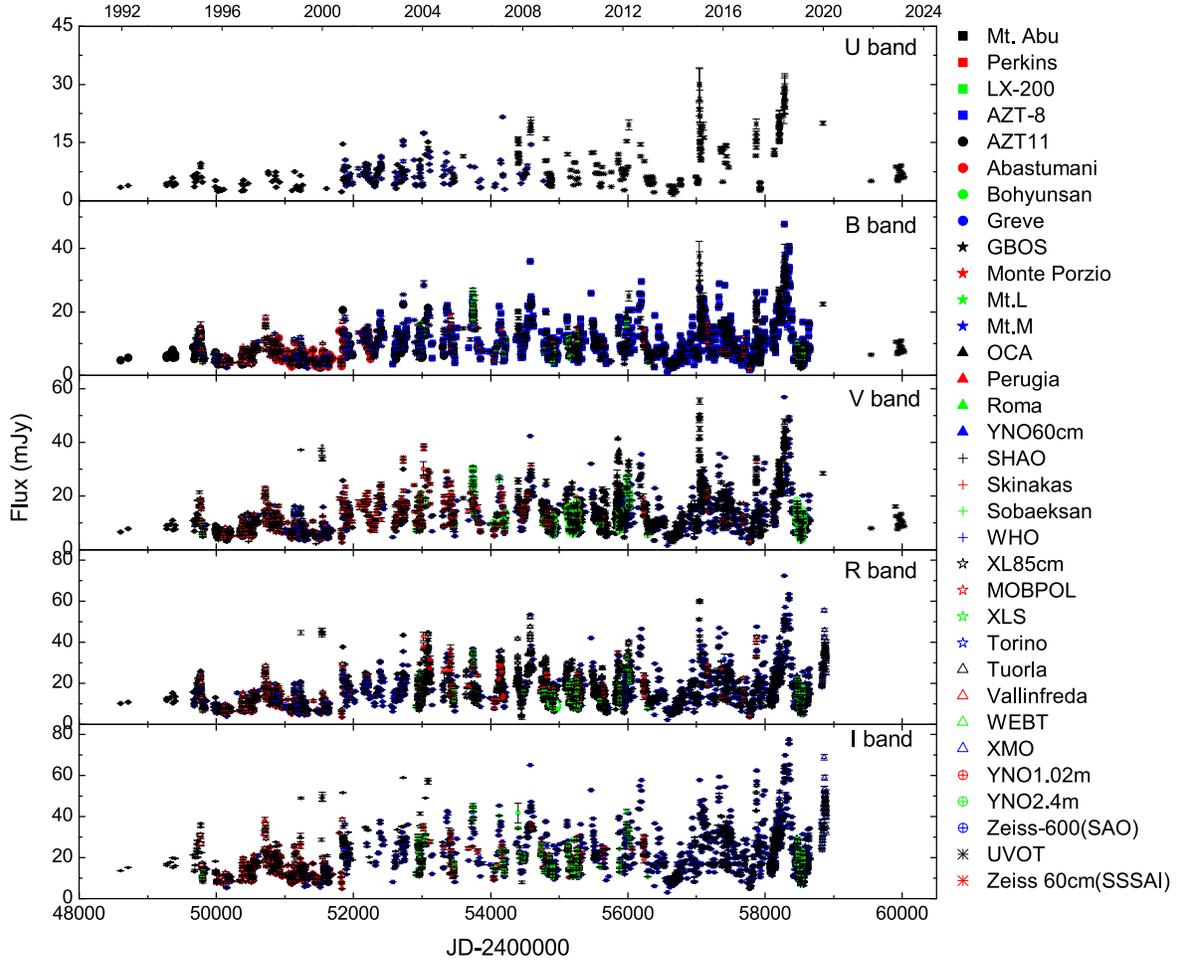

**Figure 3.** The long-term light curves of S5 0716+714 in the UBVRI bands. The symbols and colors depicted in the chart correspond to the respective sources of the data. (1) Mt. Abu: 1.2 m Telescope at Mt. Abu Infrared Observatory. (2) Perkins: 1.8 m Perkins telescope of Lowell Observatory with PRISM camera. (3) LX-200: 40 cm LX-200 telescope in St. Petersburg. (4) AZT-8: 70 cm AZT-8 telescope of the Crimean Astrophysical Observatory of the Russian Academy of Sciences. (5) AZT11: 125 cm Crimean Astrophysical Observatory telescope. (6) Abastumani: The Abastumani Observatory. (7) Bohyunsan: 1.8 m telescope of Bohyunsan Optical Astronomy Observatory. (8) Greve: The telescope of Greve. (9) GBOS: Ground-based Observational Support of the Fermi Gamma-ray Space Telescope at the University of Arizona. (10) Monte Porzio: The 70 cm f/8.3 TRC70 telescope of Monte Porzio. (11) MT.L: The 1.0 m robotic telescope of Mount Lemmon Optical Astronomy Observatory. (12) Mt.M: The Mount Maidanak Observatory. (13) OCA: Observatoire de la Côte d'Azur. (14) Perugia: the 40 cm Automatic Imaging Telescope (AIT) of the Perugia University. (15) Roma: The University of Roma group. (16) YNO60cm: The 60 cm BOOTES-4 autotelescope at the Lijiang Observatory of the Yunnan Observatories. (17) SHAO: The 1.56 m telescope at the Sheshan station of the Shanghai Astronomical Observatory. (18) Skinakas: The 1.3 m telescope of the Skinakas Observatory. (19) Sobaeksan: The 0.6 m telescope of Sobaeksan Optical Astronomy Observatory. (20) WHO: The 1.0 m telescope at Weihai Observatory. (21) XL85cm: The 85 cm telescope located at the Xinglong Station. (22) MOBPOL: The MOBPOL program. (23) XLS: The Schmidt telescope at the Xinglong Station. (24) Torino: The 1.05 m telescope of the Torino Observatory. (25) Tuorla: Tuorla Observatory blazar monitoring optical light curves. (26) Vallinfreda: The 50 cm telescope of the Vallinfreda Station. (27) WEBT: The international Whole Earth Blazar Telescope. (28) XMO: The CSP telescope at the Xingming Observatory. (29) YNO1.02 m: The 1.02 m telescope of Yunnan Astronomical Observatory. (30) YNO2.4 m: The 2.4 m telescopes at the Lijiang station of the Yunnan Astronomical Observatories. (31) Zeiss-600(SAO): Zeiss-600 telescopes of the Special Astrophysical Observatory (SAO). (32) UVOT: The Swift's Ultraviolet/Optical Telescope. (33) Zeiss 60-cm(SSSAI): the Zeiss 60 cm telescope at the Southern Station of the Sternberg Astronomical Institute.

### 3.3. Variability Amplitude

The variability amplitude (Amp) of the light curve for each night is computed using the subsequent equation to further quantify the observed variations (J. Heidt & S. J. Wagner 1996):

$$\text{Amp} = 100 \times \sqrt{(A_{\max} - A_{\min}) - 2\sigma^2_{\text{rms}}} \%, \quad (8)$$

where $A_{\max}$ and $A_{\min}$ are maximum and minimum magnitude, respectively. The rms error $\sigma$ is calculated as follows:

$$\sigma_{\text{rms}} = \sqrt{\sum \frac{(m_i - \bar{m})^2}{N - 1}}, \quad i = 1, 2, 3, \ldots, N, \quad (9)$$

where $m_i = (m_{S_1} - m_{S_2})_i$ and $\bar{m} = \overline{m_{S_1} - m_{S_2}}$ represent the differential magnitudes of stars $S_1$ and $S_2$, respectively, while the average value of $m_i$ corresponds to a specific data set for a given night.

A blazar is considered variable when the light curve on a given night passes both tests. If the light curve only satisfies one test, the blazar is regarded as a potential variable. Nonvariability of the source is established when none of the criteria are met. Typically, for a blazar to be classified as

Table 3
Statistic Parameters of Intranight Light Curves

| Date | Band | N | $\Delta T(h)$ | $F_{PeF}$ | $F_{PeF,c}$ | $F_{NA}$ | $F_{NA,c}$ | V/N | Amp. (%) | Avg. (mag) |
|---|---|---|---|---|---|---|---|---|---|---|
| (1) | (2) | (3) | (4) | (5) | (6) | (7) | (8) | (9) | (10) | (11) |
| 2018-12-22 | I | 60 | 11.58 | 0.59 | 1.66 | 1.67 | 3.58 | N | 30.59 | 13.05 |
| 2018-12-22 | R | 60 | 11.72 | 1.65 | 1.67 | 2.85 | 3.70 | N | 41.77 | 13.35 |
| 2018-12-23 | I | 41 | 3.78 | 1.14 | 1.85 | 0.91 | 4.67 | N | 25.52 | 12.87 |
| 2018-12-23 | R | 42 | 9.64 | 2.24 | 1.82 | 3.30 | 4.60 | PV | 35.60 | 13.15 |

**Note.** The initial four rows are presented herein, while the comprehensive table is available in a machine-readable format.

(This table is available in its entirety in machine-readable form in the online article.)

variable in nature, it should exhibit a variability amplitude exceeding 7.5% (Amp > 7.5%) (Y. H. Yuan & J. H. Fan 2021).

## 4. Results and Discussion

### 4.1. IDV

The analysis results for 158 light curves are presented in Table 3, which includes (1) observation date, (2) observation band, (3) number of observations per night, (4) duration in hours, (5) power-enhanced F-test value, (6) power-enhanced F-test critical value at 99% confidence, (7) nested ANOVA F-value, (8) nested ANOVA critical value at 99% confidence, (9) microvariability label (V, PV, N for variable, possible variable, and nonvariable, respectively), (10) amplitude (Amp), and (11) average magnitude. Table 3 shows the detection of IDV and possible IDV on 35 and 22 nights for the $r'$ band, respectively, and on 31 and 20 nights for the $i'$ band, respectively. IDV was detected in both bands on 26 nights, as depicted in Figures 4 and 5. Variability amplitude exceeds 7.5% (Amp > 7.5%) for both IDV and PIDV. Specific IDV details are in Table 4, including (1) observation date, (2) timescale $\Delta t$, (3) magnitude change $\Delta m$ (negative for brightness decrease), and (4) observation band. Figures 4, 5, and Table 4 highlight variability on hourly timescales during IDV nights. Notable changes include a 0.26 and 0.263 mag decrease in the $i'$ and $r'$ bands on 2019 December 10, and a 0.145 and 0.163 mag increase on 2019 December 18. On 2019 December 24, the $i'$ band showed a 0.121 mag decrease in 88.45 minutes, followed by a 0.218 mag increase over 466.38 minutes, with similar changes in the $r'$ band. On 2019 December 27, the $i'$ band exhibited a rapid 0.11 mag increase in 22.05 minutes, a 0.119 mag decrease over 165.33 minutes, and a 0.252 mag increase over 354.8 minutes, with similar changes in the $r'$ band. Further observations in the $i'$ band include a 0.219 mag decrease over 406.28 minutes on 2019 December 30, a 0.148 mag decrease over 284.55 minutes on 2020 January 15, a 0.134 mag increase over 578.27 minutes on 2020 January 18, a 0.131 mag decrease over 443.27 minutes on 2020 January 19, and a 0.146 mag increase over 426.48 minutes on 2020 January 26. Similar variations were observed in the $r'$ band. Additionally, in the $i'$ band, its brightness exhibited brightening, dimming, and brightening on 2019 December 27, 2020 January 29, and 2020 February 2. It also displayed a sequence of brightening followed by dimming on 2019 December 8 and 17, 2020 January 24, and 2020 February 1 and 3. Furthermore, it underwent a pattern of dimming succeeded by brightening on 2019 December 24, 2020 January 31, and 2020 February 13. The brightness displayed a similar variation in the $r'$ band as it did in the $i'$ band. Optical IDV of S5 0716+714 has been extensively investigated, showing high variability during 58%−85% of observation periods (e.g., S. J. Wagner & A. Witzel 1995; S. J. Wagner et al. 1996; M. Villata et al. 2000; J. Wu et al. 2005; J. Wu et al. 2012; H. Poon et al. 2009; S. Hong et al. 2017; H. T. Liu et al. 2019; J. Xu et al. 2019; D. Xiong et al. 2020; T. Tripathi et al. 2024).

The IDV behavior may be linked to relativistic jet activities or accretion disk instability (B. Rani et al. 2011; D. Xiong et al. 2017; H.-Z. Li et al. 2024a). In BL Lacertae objects, the nonthermal radiation emitted from the relativistic jet typically dominates over the thermal radiation emitted from the accretion disk. Hence, the observed IDV behavior of S5 0716+714 can be elucidated within the framework of a relativistic jet scenario, which typically assumes the propagation of a relativistic shock within a jet (R. D. Blandford & A. Königl 1979; A. P. Marscher & W. K. Gear 1985; D. Xiong et al. 2017; H.-Z. Li et al. 2024a). During the propagation of relativistic shock waves within a jet, it may interact with irregular plasma, leading to modifications in intrinsic radiation. Furthermore, the propagation direction of relativistic shocks can undergo modifications during the process, resulting in variations in the observed radiation intensity. The change in the direction of shock wave propagation results in a corresponding change of the viewing angle $\theta$, which consequently affects the Doppler factor $\delta$. The relationship between the Doppler factor and the viewing angle can be described by the following formula:

$$\delta = [\Gamma(1 - \beta \cos \theta)]^{-1}, \quad (10)$$

where $\Gamma$ and $\beta = v/c$ represent the bulk Lorentz factor and constant bulk velocity of the flow, respectively. The variability of $\delta$ would cause a change in flux $F_\nu$, with $F_\nu \propto \delta^3$.

Moreover, the optical IDV can be interpreted in a scenario of a two-fluid model plus the Kelvin–Helmholtz instability (J. Fan et al. 2023). The two fluids consist of a nonrelativistic jet comprising an electron–proton plasma, and a relativistic jet consisting of an electron–positron plasma (H. Sol et al. 1989). At the junction of the two jet components, the Kelvin–Helmholtz instability arises, giving rise to significant disturbances (G. E. Romero 1995; J. T. Cai et al. 2022). The instability has been observed within the emission region of a substantial number of Fermi blazars. The emission would exhibit detectable variability when the instability is amplified due to accumulation. The timescale of variability is contingent upon the efficiency of energy dissipation. Additionally, the emergence of detectable violent variability may occur if the instability manifests as kink instabilities, which have the

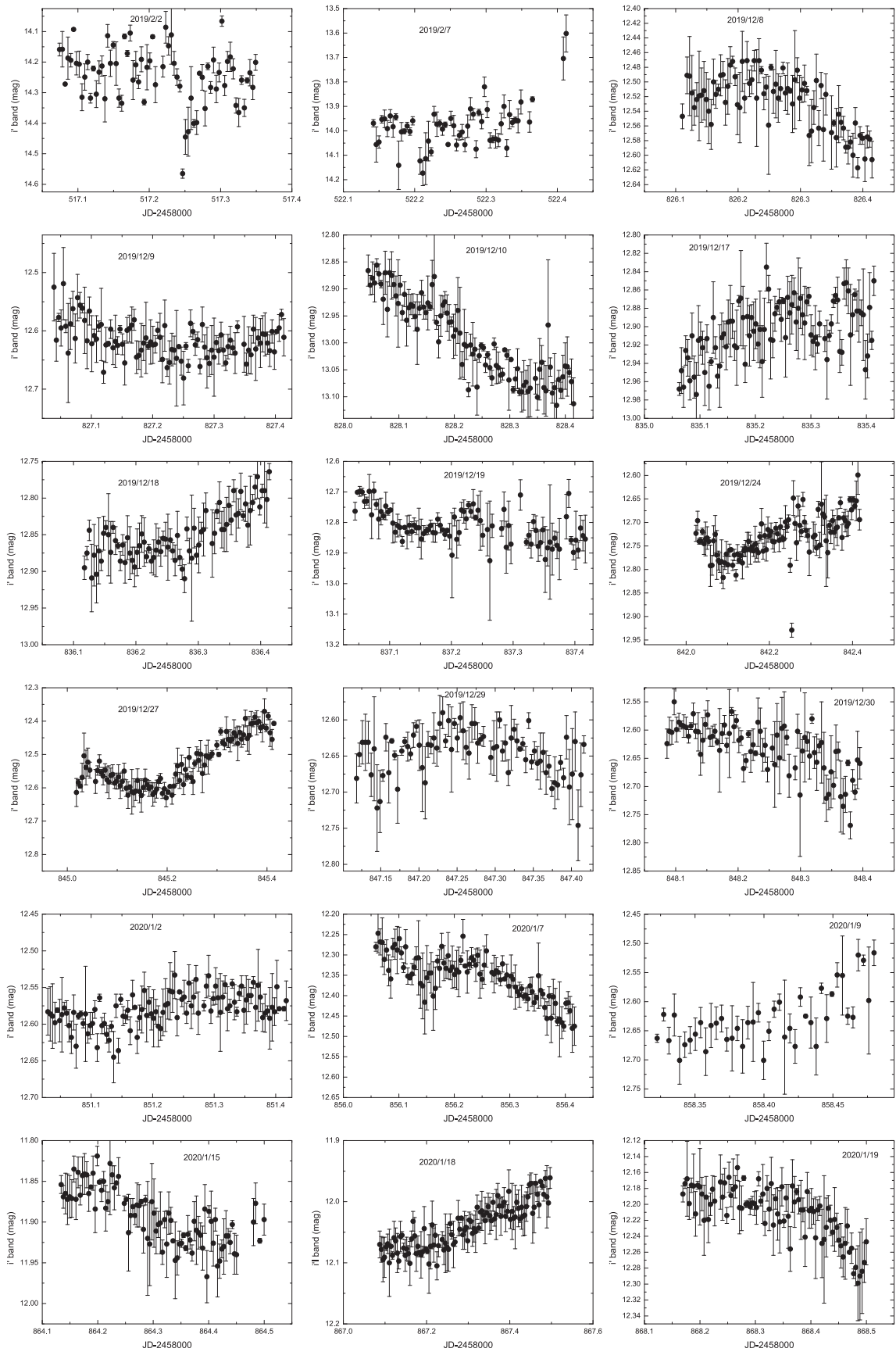

**Figure 4.** $i'$-band intraday variable light curves of S5 0716+714.

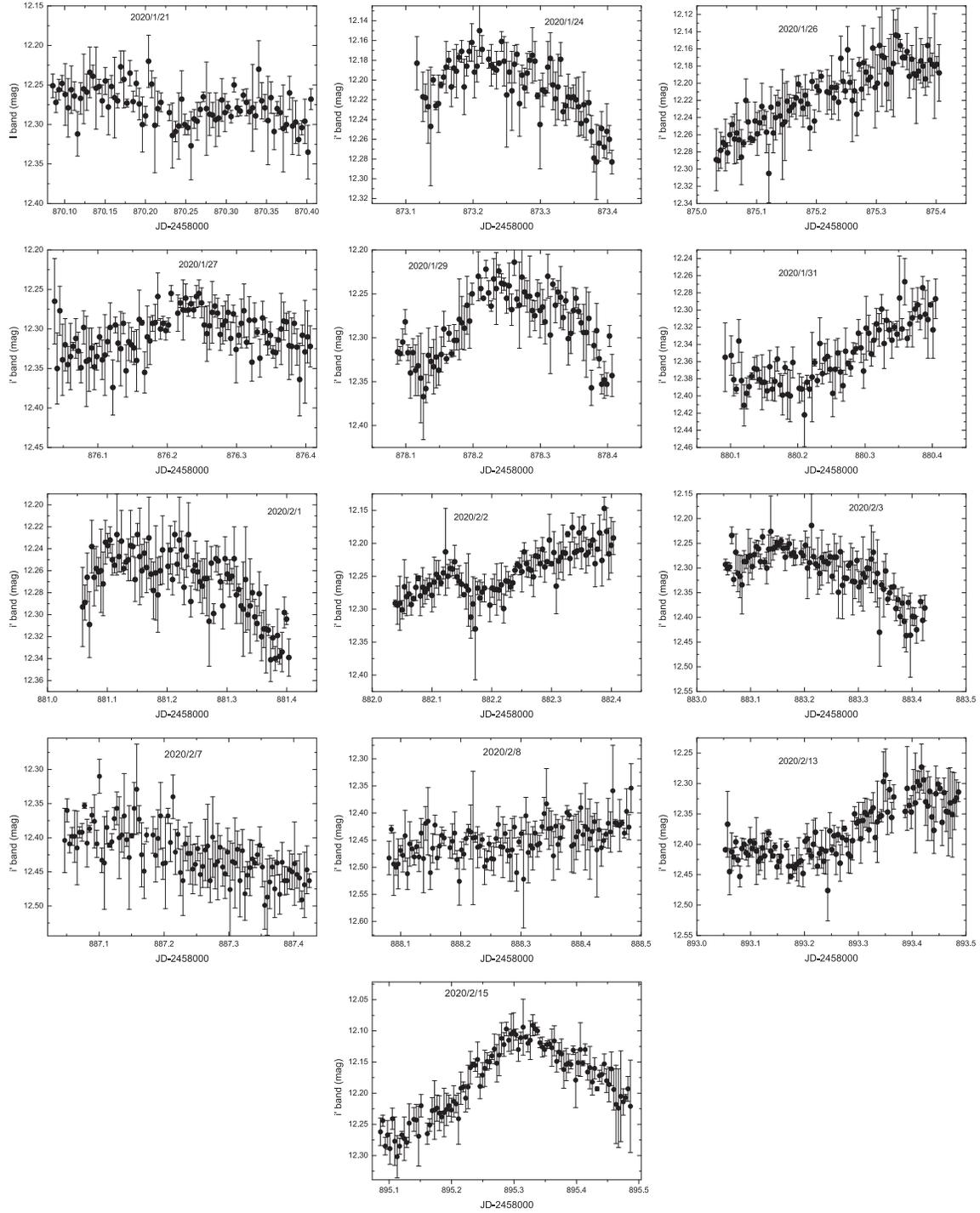

**Figure 4.** (Continued.)

potential to disrupt the jet and initiate magnetic reconnection (e.g., D. Giannios & H. C. Spruit 2006; O. Porth & S. S. Komissarov 2015; R. Barniol Duran et al. 2017; A. Shukla & K. Mannheim 2020; J. Fan et al. 2023). This process of magnetic reconnection facilitates particle acceleration, leading to rapid dissipation of these particles within an extremely short temporal scale (A. Shukla & K. Mannheim 2020; J. Fan et al. 2023). The occurrence of Kelvin–Helmholtz instability is contingent upon the strength of the magnetic field, and it only manifests when the magnetic field strength falls below a critical value $B_c$ (J. T. Cai et al. 2022; J. Fan et al. 2023; T. Tripathi et al. 2024), which can be determined by employing the following formula (G. E. Romero 1995):

$$B_c = [4\pi n_e m_e c^2(\gamma^2 - 1)]^{1/2}\gamma^{-1}, \quad (11)$$

where $n_e$, $m_e$, and $c$ are the local electron density, the rest mass of electrons, and speed of light, respectively. When the axial magnetic field exceeds the critical value $B_c$, it can effectively suppress the Kelvin–Helmholtz instabilities and bends at the base of the jet, which can cause the IDV behavior of blazars

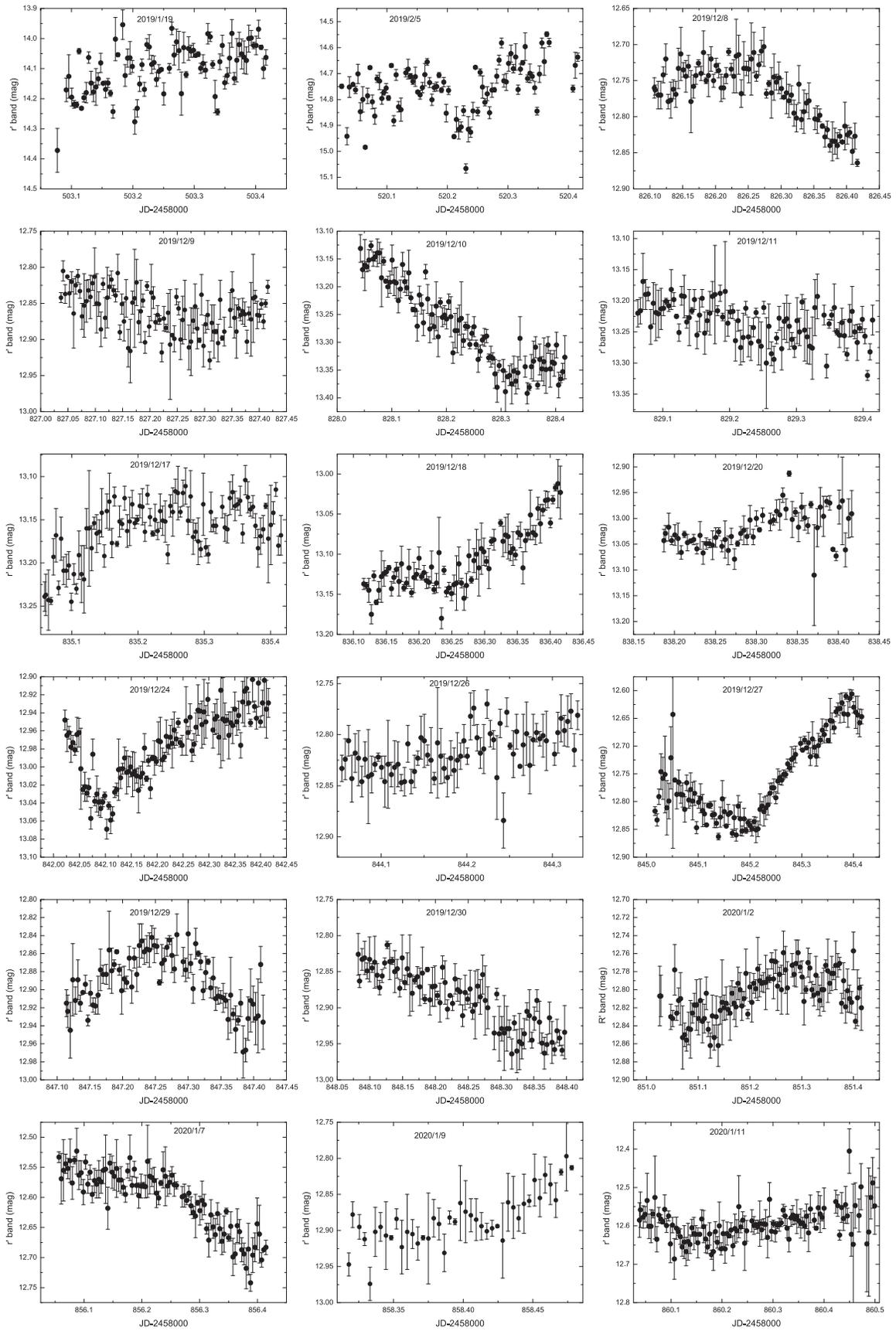

**Figure 5.** $r'$-band intraday variable light curves of S5 0716+714.

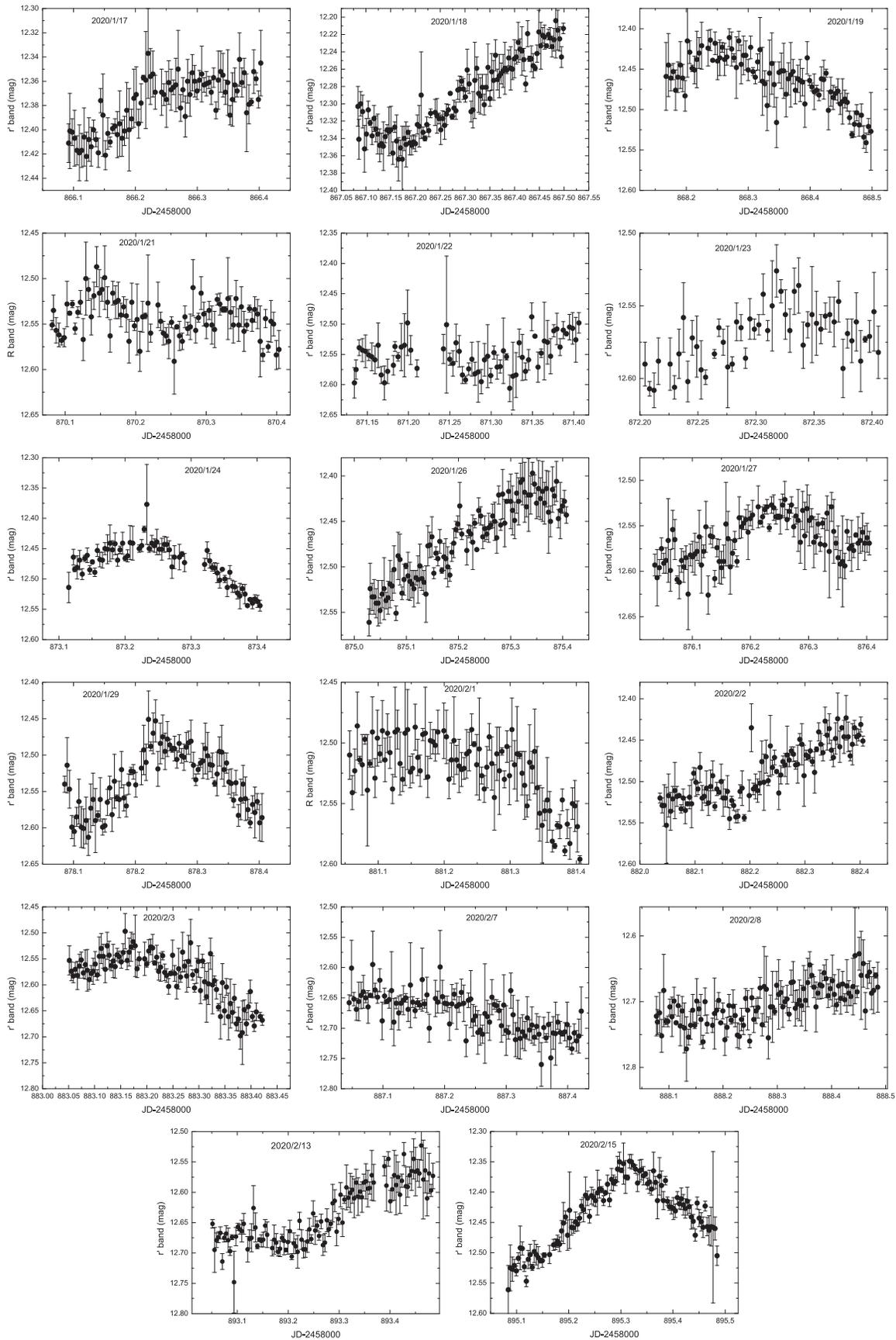

**Figure 5.** (Continued.)

Table 4
The Variability of IDV Night

| Date (1) | Δt (min) (2) | Δm[a] (3) | Band (4) | Date (1) | Δt (min) (2) | Δm[a] (3) | Band (4) | Date (1) | Δt (min) (2) | Δm[a] (3) | Band (4) |
|---|---|---|---|---|---|---|---|---|---|---|---|
|  | 220.10 | −0.472 |  |  | 38.52 | −0.085 |  | 2019-12-29 | 179.18 | 0.103 | $r'$ |
| 2019-02-02 | 79.33 | 0.499 | $i'$ | 2020-01-29 | 134.17 | 0.145 | $i'$ |  | 120.77 | −0.131 |  |
|  | 34.00 | −0.299 |  |  | 258.63 | −0.131 |  | 2019-12-30 | 447.77 | −0.133 | $r'$ |
| 2019-02-07 | 65.75 | −0.234 | $i'$ | 2020-01-31 | 140.27 | −0.086 | $i'$ |  | 55.07 | 0.052 |  |
|  | 289.32 | 0.571 |  |  | 214.58 | 0.155 |  | 2020-01-02 | 11.02 | −0.058 | $r'$ |
| 2019-12-08 | 99.57 | 0.087 | $i'$ |  | 65.93 | 0.082 |  |  | 244.25 | 0.103 |  |
|  | 238.72 | −0.146 |  | 2020-02-01 | 98.87 | −0.055 | $i'$ | 2020-01-07 | 432.80 | −0.219 | $r'$ |
| 2019-12-09 | 143.88 | −0.136 | $i'$ |  | 274.58 | −0.112 |  | 2020-01-09 | 43.87 | −0.052 | $r'$ |
| 2019-12-10 | 466.85 | −0.260 | $i'$ |  | 104.68 | 0.088 |  |  | 70.17 | 0.117 |  |
| 2019-12-12 | 411.18 | −0.142 | $i'$ | 2020-02-02 | 71.57 | −0.117 | $i'$ | 2020-01-11 | 88.18 | −0.141 | $r'$ |
| 2019-12-17 | 115.80 | 0.105 | $i'$ |  | 310.40 | 0.183 |  |  | 527.00 | 0.179 |  |
|  | 157.15 | −0.101 |  | 2020-02-03 | 77.07 | 0.108 |  | 2020-01-17 | 99.23 | 0.084 | $r'$ |
| 2019-12-18 | 409.83 | 0.145 | $i'$ |  | 362.27 | −0.205 |  |  | 42.47 | −0.038 |  |
|  | 121.36 | −0.156 |  | 2020-02-07 | 159.80 | −0.142 | $i'$ | 2020-01-18 | 120.80 | −0.064 | $r'$ |
| 2019-12-19 | 49.65 | 0.168 | $i'$ | 2020-02-13 | 168.90 | −0.086 | $i'$ |  | 446.53 | 0.160 |  |
|  | 38.62 | −0.186 |  |  | 350.07 | 0.180 |  | 2020-01-19 | 296.20 | −0.126 | $r'$ |
| 2019-12-24 | 88.45 | −0.121 | $i'$ | 2020-02-15 | 268.52 | 0.201 | $i'$ | 2020-01-21 | 88.18 | 0.267 | $r'$ |
|  | 466.38 | 0.218 |  |  | 242.07 | −0.123 |  |  | 44.07 | −0.058 |  |
|  | 22.05 | 0.110 |  |  | 22.90 | 0.289 |  | 2020-01-22 | 40.32 | 0.099 | $r'$ |
| 2019-12-27 | 165.33 | −0.119 | $i'$ | 2019-01-19 | 28.23 | −0.323 | $r'$ |  | 60.52 | −0.094 |  |
|  | 354.80 | 0.252 |  |  | 22.60 | −0.260 |  |  | 120.73 | 0.108 |  |
|  | 27.67 | −0.091 |  |  | 11.30 | 0.221 |  | 2020-01-23 | 152.00 | 0.082 | $r'$ |
| 2019-12-29 | 74.25 | 0.113 | $i'$ | 2019-02-05 | 92.07 | −0.411 | $r'$ |  | 82.43 | −0.067 |  |
|  | 16.55 | −0.078 |  |  | 84.64 | 0.484 |  | 2020-01-24 | 135.18 | 0.074 | $r'$ |
|  | 109.78 | −0.090 |  | 2019-12-08 | 205.43 | −0.161 | $r'$ |  | 252.52 | −0.126 |  |
| 2019-12-30 | 406.28 | −0.219 | $i'$ | 2019-12-09 | 33.18 | −0.108 | $r'$ | 2020-01-26 | 451.32 | 0.164 | $r'$ |
| 2020-01-02 | 33.05 | −0.081 | $i'$ |  | 16.58 | 0.095 |  | 2020-01-27 | 156.95 | 0.099 | $r'$ |
|  | 16.53 | 0.079 |  | 2019-12-10 | 354.27 | −0.263 | $r'$ |  | 148.63 | −0.073 |  |
|  | 121.30 | −0.169 |  | 2019-12-11 | 88.58 | −0.115 | $r'$ | 2020-01-29 | 139.65 | 0.162 | $r'$ |
| 2020-01-07 | 44.08 | 0.137 | $i'$ |  | 33.15 | −0.077 |  |  | 258.63 | −0.142 |  |
|  | 338.90 | −0.201 |  | 2019-12-17 | 93.73 | 0.122 | $r'$ | 2020-02-01 | 148.18 | −0.107 | $r'$ |
| 2020-01-09 | 203.00 | 0.185 | $i'$ |  | 71.68 | −0.073 |  | 2020-02-02 | 266.38 | 0.091 | $r'$ |
| 2020-01-15 | 284.55 | −0.148 | $i'$ | 2019-12-18 | 409.83 | 0.163 | $r'$ | 2020-02-03 | 318.18 | −0.201 | $r'$ |
| 2020-01-18 | 578.27 | 0.134 | $i'$ | 2019-12-20 | 96.05 | 0.166 | $r'$ | 2020-02-07 | 187.18 | −0.085 | $r'$ |
| 2020-01-19 | 443.27 | −0.131 | $i'$ | 2019-12-24 | 116.18 | −0.121 | $r'$ | 2020-02-08 | 327.60 | 0.128 | $r'$ |
| 2020-01-21 | 42.55 | −0.094 | $i'$ |  | 405.78 | 0.166 |  | 2020-02-13 | 366.23 | 0.183 | $r'$ |
| 2020-01-24 | 104.57 | 0.097 | $i'$ | 2019-12-26 | 44.37 | 0.068 | $r'$ | 2020-02-15 | 307.23 | 0.211 | $r'$ |
|  | 283.13 | −0.133 |  | 2019-12-27 | 209.42 | −0.114 | $r'$ |  | 269.53 | −0.155 |  |
| 2020-01-26 | 426.48 | 0.146 | $i'$ |  | 321.73 | 0.254 |  |  |  |  |  |

**Note.**
[a] The negative sign indicates a decrease in brightness.

(T. Tripathi et al. 2024). Therefore, the observed IDV behavior of S5 0716 + 714 suggests that the magnetic field strength in the jet is expected to be lower than that of the critical magnetic field ($B_c$). The value of $B_c$ has been estimated by T. Tripathi et al. (2024), who found that it should fall within the range of 0.07–0.70 G. Additionally, the magnetic field strength of S5 0716+714 has also been estimated through analysis of multiwavelength SEDs, yielding a range of 0.01–0.58 G (N. H. Liao et al. 2014; MAGIC Collaboration et al. 2018). However, it is important to note that while all blazars have the potential to generate instabilities, the occurrence of strong and violent variability is sporadic due to the majority of these instabilities diminishing before they can amplify and produce significant variations.

In addition, the presence of turbulence in the jet can cause IDV behavior. If turbulence exists, the particles within each turbulent cell would be accelerated by the relativistic shock and cooled through synchrotron emission. During this process, the emission flux would increase, and there would be variations in the density, size, and magnetic field direction of each individual turbulent cell. However, the radiation flux would attenuate following the passage of the shock through each turbulent cell (T. Tripathi et al. 2024).

### 4.2. The Minimum Variability Timescale and Black Hole Mass Estimation

In general, IDV can be generated in close proximity to the central supermassive black hole. Therefore, the minimum variability timescale can be utilized to estimate the size of the radiation region and the masses of the black hole (H. R. Miller et al. 1989; D. Xiong et al. 2017; H. T. Liu & J. M. Bai 2015; X. Chang et al. 2023). The minimum variability timescale can be estimated through the autocorrelation function (ACF) analysis (T. Alexander 1997). The ACF can be computed using the following formula (D. Xiong et al. 2017; X. Chang et al. 2023),

$$\text{ACF}(\tau) = \overline{(m(t) - \overline{m}) \cdot (m(t + \tau) - \overline{m})}, \quad (12)$$

where $m$ and $\tau$ are the observed magnitude and time lags, respectively. Moreover, the top line represents a temporal average, indicating the mean value over time.

The ACF can quantitatively assess the temporal correlation of the light curve by evaluating its time shifted as a function of the time lags $\tau$. The zero-crossing time of the ACF, which serves as a characteristic timescale for variability, is defined as the minimum timescale required for the ACF to reach zero (T. Alexander 1997; H. T. Liu et al. 2008; D. Xiong et al. 2017). The width of the ACF peak near zero time lag is directly proportional to the typical timescale of an underlying signal observed in the light curve (T. Alexander 1997; U. Giveon et al. 1999; H. T. Liu et al. 2008; D. Xiong et al. 2017). Thus, the zero-crossing time of ACF can serve as a temporal scale for variability. Furthermore, we employed Monte Carlo simulations (e.g., X. Yang et al. 2020; Y.-F. Wang & Y.-G. Jiang 2021; H.-Z. Li et al. 2024b) to estimate the uncertainties associated with the minimum timescale. Initially, 10,000 artificial light curves were generated and resampled based on the statistical parameters and the irregularity sampling effect derived from the observed data. Additionally, the corresponding minimum timescales were computed using the autocorrelation function method. Subsequently, we calculated the standard deviation of these minimum timescales from the 10,000 artificial light curves as an estimate for quantifying errors in determining the minimum timescale of our observed data. The IDV observation data obtained from 31 nights of the $i'$ band and 35 nights of the $r'$ band will be analyzed using the ACF method to search for the minimum timescale of variability.

To confirm the zero-crossing time, the results of ACF were fitted by the ninth-order polynomial least squares (U. Giveon et al. 1999). Figures 6 and 7, along with Table 5, present the results of the ACF calculations. These results reveal eight relatively short timescales, corresponding to the $r'$-band observations on 2019 January 19 and December 11, with timescales of $9.59 \pm 0.93$ minutes and $7.98 \pm 1.66$ minutes, respectively. The $i'$ band on 2020 January 9 and 18, with timescales of $6.49 \pm 13.11$ and $8.97 \pm 25.64$ minutes, respectively. Moreover, both the $i'$ and $r'$ bands on 2020 February 7 and 8 show very short timescales. On 2020 February 7, the timescales for the $i'$ and $r'$ bands were $7.33 \pm 0.74$ and $8.54 \pm 7.54$ minutes, respectively, while on February 8, they were $8.96 \pm 8.96$ and $10.20 \pm 14.26$ minutes, respectively. However, apart from the timescales of the $r'$ band on 2019 January 9 and the $i'$ band on 2020 February 7, which exhibit relatively small errors, the errors associated with the other timescales are significantly larger. Consequently, it becomes evident that these minimum timescales lack reliability. Therefore, the reliability minimum timescale of variability should be $7.33 \pm 0.74$ minutes for the $i'$ band on 2020 February 7. Similar results were reported for this source in several literature (e.g., H. Poon et al. 2009; B. K. Zhang et al. 2012; B.-z. Dai et al. 2015; D. Xiong et al. 2017; S. Hong et al. 2018; H. T. Liu et al. 2019; T. Tripathi et al. 2024) with the timescale in the range of 0.1–2.8 hr. Considering the variability timescales as the light-crossing time of the emitting blob, the minimum variability timescale can serve as an upper limit for determining the sizes of emission regions. The range of the emission region in the jet can be determined by the following equation:

$$R \leqslant \frac{c\delta\Delta t_{\min}}{1 + z}, \quad (13)$$

where $c$, $\delta$, $\Delta t_{\min}$, and $z$ represent the speed of light, Doppler factor, minimum variability timescale in seconds, and redshift, respectively. The Doppler factor $\delta$ estimation of S5 0716+714 is within the range of 4 to 30, namely $4 \leqslant \delta \leqslant 30$ (U. Bach et al. 2005; R. Nesci et al. 2005; T. Hovatta et al. 2009; S. G. Jorstad et al. 2017; E. V. Kravchenko et al. 2020). According to Equation (13), the upper limit of $R$ is in the range of $(0.40 \sim 3.02) \times 10^{14}$ cm, with $\Delta t_{\min} = 7.33 \pm 0.74$ minutes and $z = 0.31$. Additionally, the minimum timescales of variability can also serve as estimations for the upper limit of the mass of a black hole located at the center of blazars (e.g., G. Z. Xie et al. 2002; A. C. Gupta et al. 2009; H. T. Liu & J. M. Bai 2015). Based on the model proposed by H. T. Liu & J. M. Bai (2015), the upper limit of the mass of a black hole $M_{\text{BH}}$ can be obtained by calculation using the following formulae:

$$M_{\text{BH}} \leqslant 1.70 \times 10^4 \frac{\delta \Delta t_{\min}}{1 + z} M_\odot (j = 0), \quad (14)$$

and

$$M_{\text{BH}} \leqslant 5.09 \times 10^4 \frac{\delta \Delta t_{\min}}{1 + z} M_\odot (j \sim 1), \quad (15)$$

where $\delta$, $\Delta t_{\min}$, $z$, and $M_\odot$ is the Doppler factor, minimum variability timescale in seconds, redshift, and the mass of the Sun, respectively. $j$ characterizes the dimensionless spin parameter of the black hole, with $j = 0$ corresponding to Schwarzschild black holes and $j \sim 1$ representing Kerr black holes. The central black hole in blazars is generally believed to be a rotating Kerr black hole, as suggested by G.-Z. Xie et al. (2005). Thus, the upper limit of the black hole masses $M_{\text{BH}}$ in S5 0716+714 is estimated to be about $0.68 \sim 5.12 \times 10^8 M_\odot$, as calculated using Equation (15) with parameters $4 \leqslant \delta \leqslant 30$ (U. Bach et al. 2005; R. Nesci et al. 2005; T. Hovatta et al. 2009; S. G. Jorstad et al. 2017; E. V. Kravchenko et al. 2020), $\Delta t_{\min} = 7.33 \pm 0.74$ minutes, and $z = 0.31$. The SMBH mass in S5 0716+714 has been estimated by numerous researchers, yet its determination remains elusive (D. Ł. Król et al. 2023). The masses obtained by us are consistent with the result reported in the literature. The black hole masses of S5 0716 +714 were estimated to be $10^{7.87} M_\odot$ by B. K. Zhang et al. (2012). X.-L. Liu et al. (2021) reported reported a range of black hole masses for this source, from $4.51 \times 10^7 M_\odot$ to $2.74 \times 10^8 M_\odot$. Y.-H. Yuan et al. (2017) found that the masses fall within the range of $(4.2 \sim 25.6) \times 10^7 M_\odot$. M. S. Butuzova (2021)'s study yielded a mass estimate of $5.0 \times 10^8 M_\odot$ for this source. G. Bhatta et al. (2016b) reported a range of masses for this source, from $3.0 \times 10^8 M_\odot$ to $4.0 \times 10^9 M_\odot$. Additionally, H. T. Liu et al. (2019), N. Kaur et al. (2018), E. W. Liang & H. T. Liu (2003), and A. Agarwal et al. (2016) reported black holes with masses of $8.13 \times 10^8$, $5.6 \times 10^8$, $1.25 \times 10^8$, and $2.42 \times 10^9 M_\odot$, respectively. Furthermore, A. C. Gupta et al. (2009) and S. Hong et al. (2018) estimated the masses as being equal to or less than $2.5 \times 10^6$ and $5 \times 10^6$, respectively. The study of B.-z. Dai et al. (2015) found that the mass estimates ranged from $1.1 \times 10^{10} M_\odot$ to $4.4 \times 10^{10} M_\odot$.

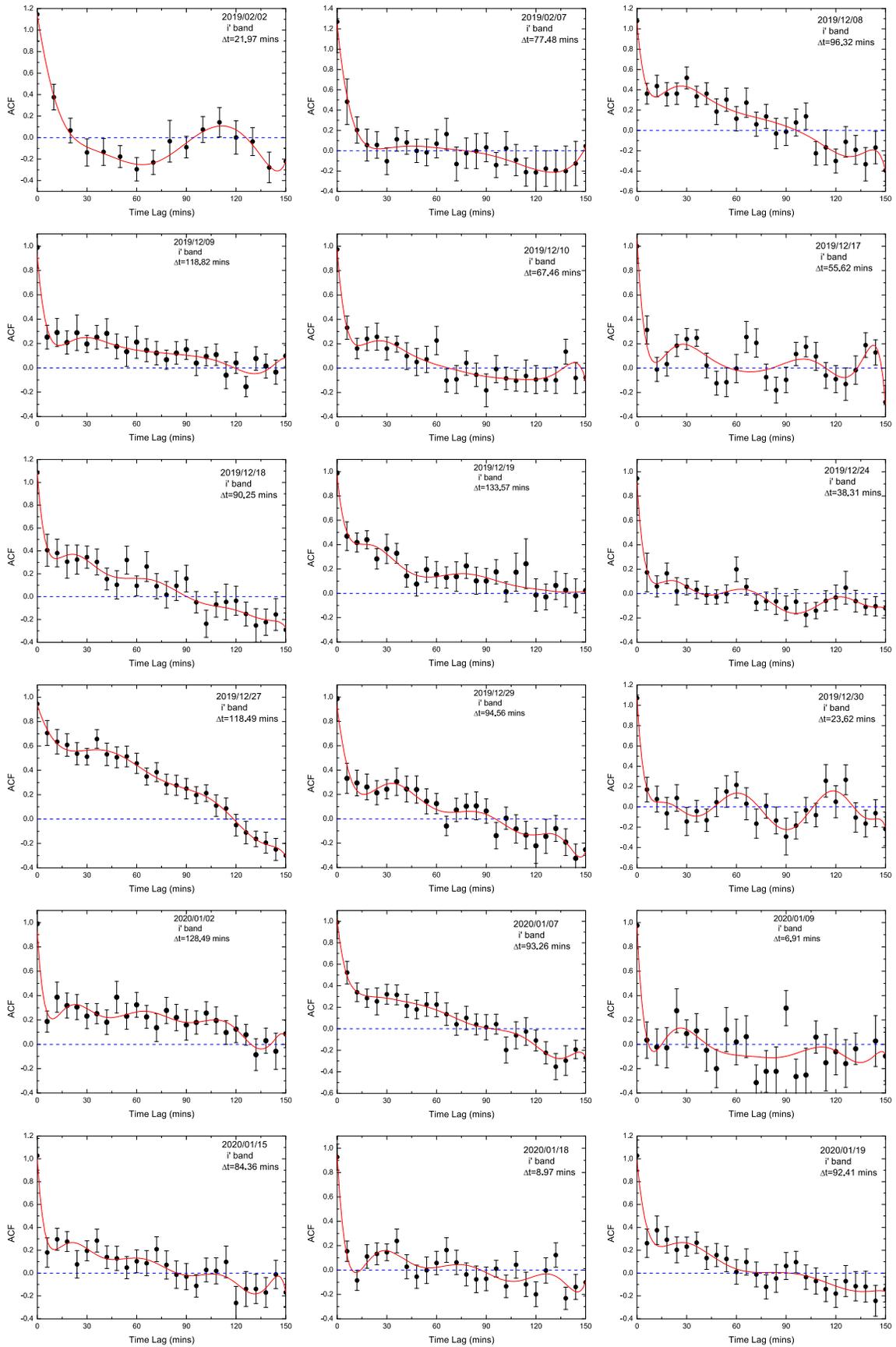

**Figure 6.** Results of the ACF analysis for $i'$-band IDV light curves. The red line is a polynomial least-squares fitting.

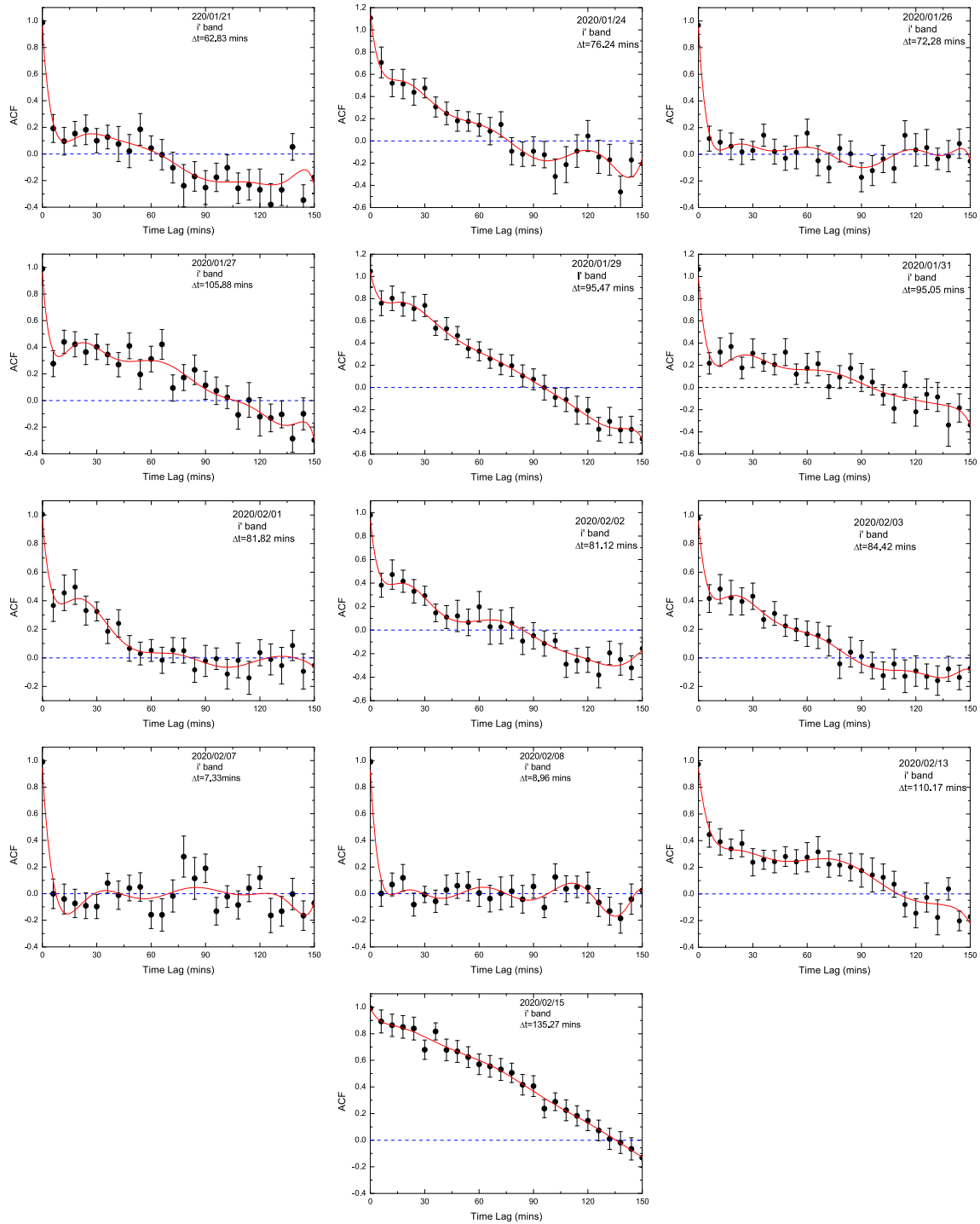

**Figure 6.** (Continued.)

### 4.3. The Color Behavior

The color behavior can generally be observed in blazars, which facilitates the study of their emission properties. Different spectral behaviors, indicating distinct radiation processes, have been reported in the literature for S5 0716 +714, with variations observed on short timescales. The source has exhibited both BWB and RWB behaviors at different instances, while sometimes showing no spectral variability (e.g., G. Ghisellini et al. 1997; C. M. Raiteri et al. 2003; J. Wu et al. 2005, 2007; C. S. Stalin et al. 2006; H. Poon et al. 2009; S. Hong et al. 2017; D. Xiong et al. 2017; C.-J. Wang et al. 2019; T. Tripathi et al. 2024). Hence, we conducted an analysis of the color behaviors of S5 0716+714 in order to investigate its emission properties. The color index ($r'-i'$) was calculated by utilizing the quasi-simultaneous observation data obtained from the $i'$ and $r'$ wave bands. The correlation between the $r'-i'$ color index and the $r'$-band magnitude is investigated using an error-weighted linear regression analysis. The color–magnitude diagrams for each individual night were plotted in Figure 8, while the results of the error-weighted linear regression analysis can be found in Table 6. In linear regression analysis, the absolute value of the correlation coefficient ($r$) falling within 0.5–1.0, 0.3–0.5, 0.1–0.3, and 0–0.1 ranges indicates strong, moderate, weak, and no correlation, respectively (X. Chang et al. 2023). Moreover, the chance probability value of $p < 0.001$ indicates a statistically significant correlation, whereas a lack of correlation is observed when $p > 0.05$ (X. Chang et al. 2023). In addition, Y. Fang et al. (2022) proposed that the reliability of the color–magnitude correlation can be established when the absolute value of the correlation coefficient $r$ exceeds 0.2 and the chance probability $p$ is below 0.01.

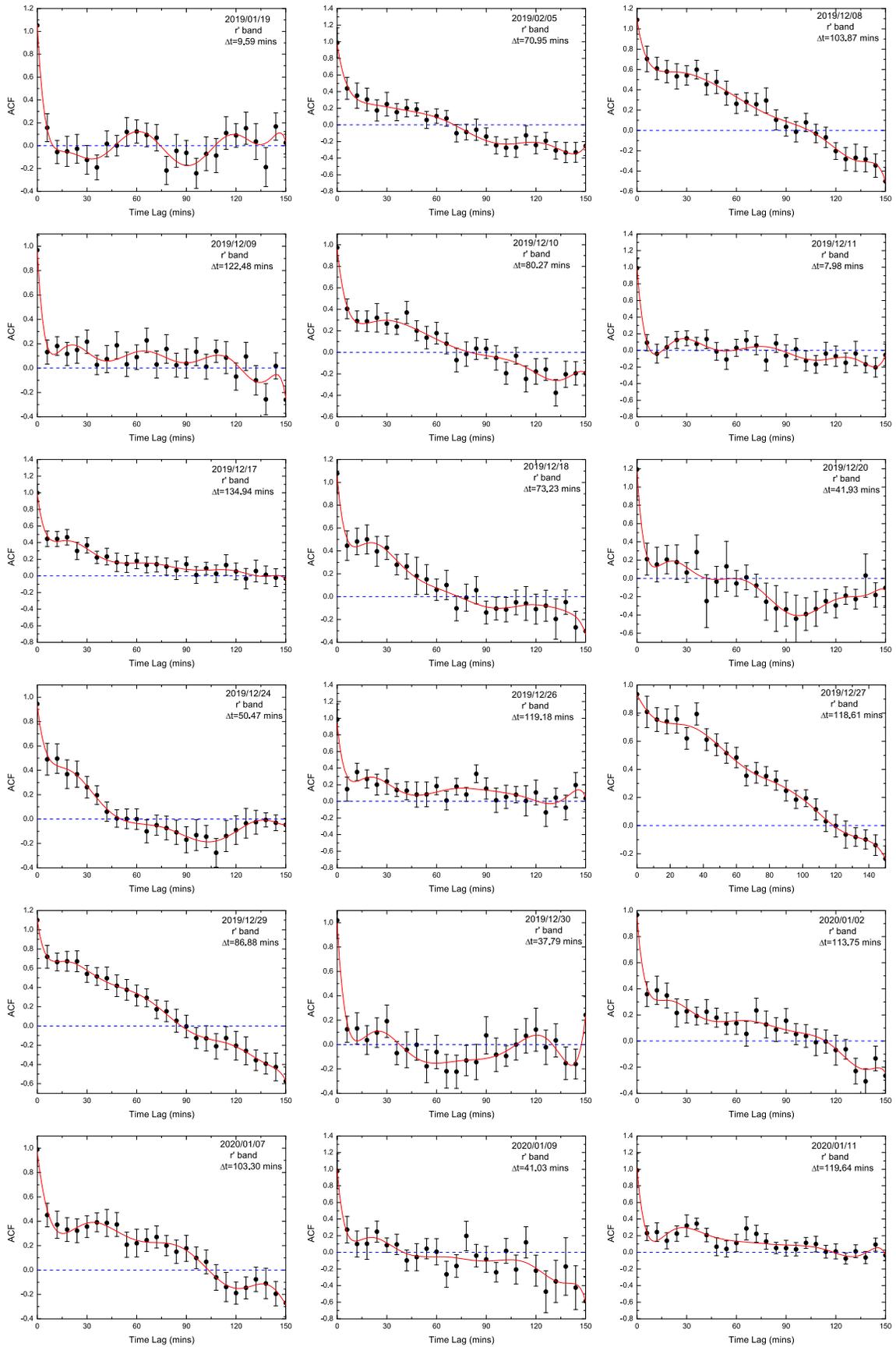

**Figure 7.** Results of the ACF analysis for $r'$-band IDV light curves. The red line is a polynomial least-squares fitting.

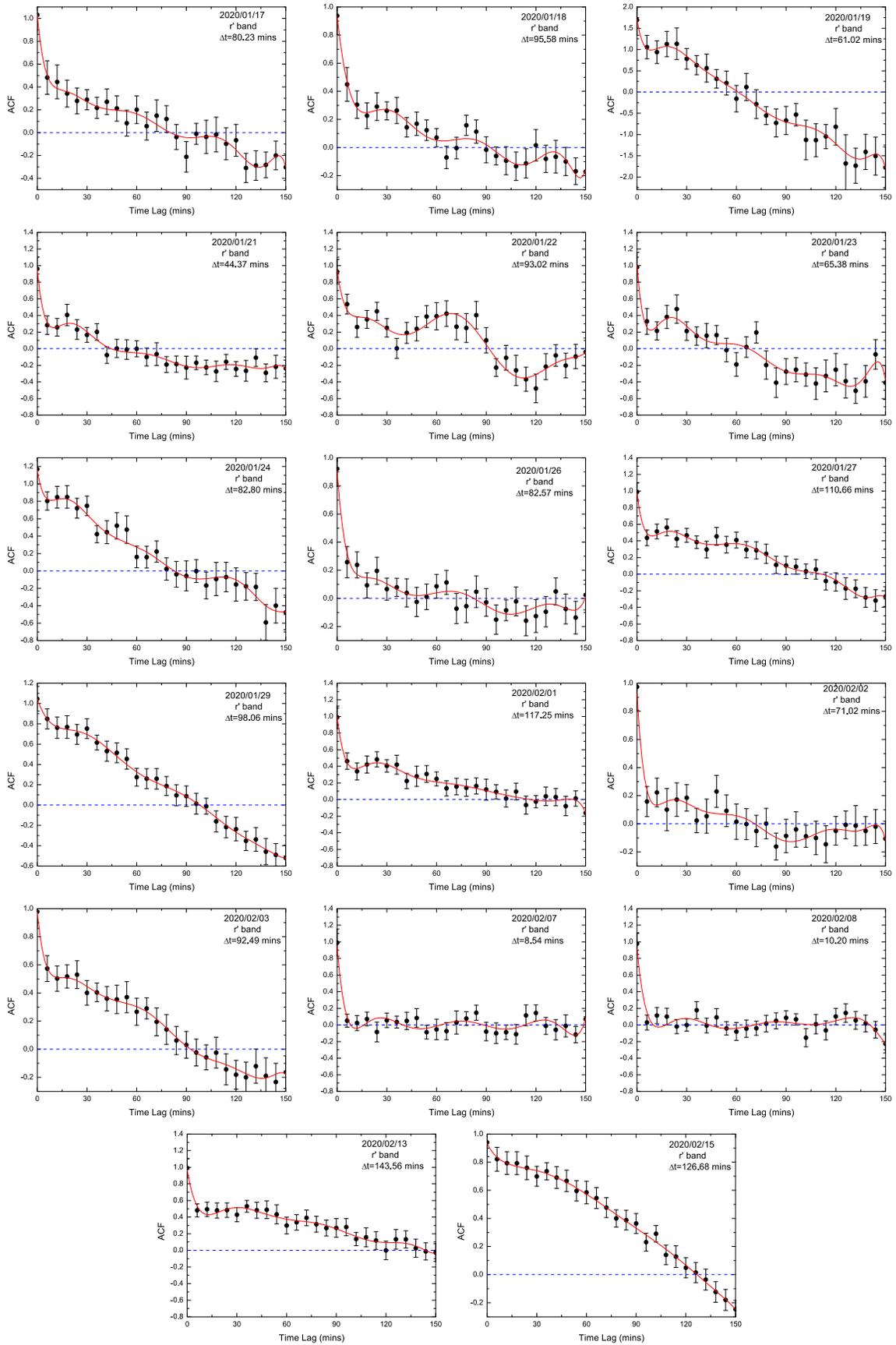

**Figure 7.** (Continued.)

Table 5
The Minimum Variability Timescale of IDV Night

| Date (1) | $\triangle t$ (min) (2) | Band (3) | Date (1) | $\triangle t$ (min) (2) | Band (3) |
|---|---|---|---|---|---|
| 2019-01-19 | 9.59 ± 0.93 | $r'$ | 2020-01-15 | 84.36 ± 12.32 | $i'$ |
| 2019-02-02 | 21.97 ± 1.22 | $i'$ | 2020-01-17 | 80.23 ± 7.55 | $r'$ |
| 2019-02-05 | 70.95 ± 0.98 | $r'$ | 2020-01-18 | 8.97 ± 25.64 | $i'$ |
| 2019-02-07 | 77.48 ± 30.76 | $i'$ | 2020-01-18 | 95.58 ± 12.12 | $r'$ |
| 2019-12-08 | 96.32 ± 4.02 | $i'$ | 2020-01-19 | 92.41 ± 16.65 | $i'$ |
| 2019-12-08 | 103.87 ± 1.67 | $r'$ | 2020-01-19 | 61.02 ± 4.10 | $r'$ |
| 2019-12-09 | 118.82 ± 7.80 | $i'$ | 2020-01-21 | 62.83 ± 18.50 | $i'$ |
| 2019-12-09 | 122.48 ± 15.69 | $r'$ | 2020-01-21 | 44.37 ± 3.24 | $r'$ |
| 2019-12-10 | 67.46 ± 7.87 | $i'$ | 2020-01-22 | 93.02 ± 1.47 | $r'$ |
| 2019-12-10 | 80.27 ± 5.57 | $r'$ | 2020-01-23 | 65.38 ± 5.28 | $r'$ |
| 2019-12-11 | 7.98 ± 11.66 | $r'$ | 2020-01-24 | 76.24 ± 2.62 | $i'$ |
| 2019-12-17 | 55.62 ± 15.18 | $i'$ | 2020-01-24 | 82.80 ± 2.18 | $r'$ |
| 2019-12-17 | 134.94 ± 8.44 | $r'$ | 2020-01-26 | 72.28 ± 26.66 | $i'$ |
| 2019-12-18 | 90.25 ± 4.50 | $i'$ | 2020-01-26 | 82.57 ± 19.43 | $r'$ |
| 2019-12-18 | 73.23 ± 5.12 | $r'$ | 2020-01-27 | 105.88 ± 5.68 | $i'$ |
| 2019-12-19 | 133.57 ± 13.05 | $i'$ | 2020-01-27 | 110.66 ± 6.14 | $r'$ |
| 2019-12-20 | 41.93 ± 12.31 | $r'$ | 2020-01-29 | 95.47 ± 1.81 | $i'$ |
| 2019-12-24 | 38.31 ± 13.76 | $i'$ | 2020-01-29 | 98.06 ± 1.20 | $r'$ |
| 2019-12-24 | 50.47 ± 4.34 | $r'$ | 2020-01-31 | 95.05 ± 4.59 | $i'$ |
| 2019-12-26 | 119.18 ± 21.92 | $r'$ | 2020-02-01 | 81.82 ± 13.99 | $i'$ |
| 2019-12-27 | 118.49 ± 1.04 | $i'$ | 2020-02-01 | 117.25 ± 12.30 | $r'$ |
| 2019-12-27 | 118.61 ± 1.30 | $r'$ | 2020-02-02 | 81.12 ± 5.40 | $i'$ |
| 2019-12-29 | 94.56 ± 2.62 | $i'$ | 2020-02-02 | 71.02 ± 6.66 | $r'$ |
| 2019-12-29 | 86.88 ± 1.28 | $r'$ | 2020-02-03 | 84.42 ± 4.62 | $i'$ |
| 2019-12-30 | 23.62 ± 10.39 | $i'$ | 2020-02-03 | 92.49 ± 2.93 | $r'$ |
| 2019-12-30 | 37.79 ± 14.90 | $r'$ | 2020-02-07 | 7.33 ± 0.74 | $i'$ |
| 2020-01-02 | 128.49 ± 3.00 | $i'$ | 2020-02-07 | 8.54 ± 7.54 | $r'$ |
| 2020-01-02 | 113.75 ± 3.78 | $r'$ | 2020-02-08 | 8.96 ± 8.96 | $i'$ |
| 2020-01-07 | 93.26 ± 9.11 | $i'$ | 2020-02-08 | 10.20 ± 14.26 | $r'$ |
| 2020-01-07 | 103.30 ± 1.49 | $r'$ | 2020-02-13 | 110.17 ± 5.26 | $i'$ |
| 2020-01-09 | 6.91 ± 13.11 | $i'$ | 2020-02-13 | 143.56 ± 2.94 | $r'$ |
| 2020-01-09 | 41.03 ± 11.64 | $r'$ | 2020-02-15 | 135.27 ± 1.49 | $i'$ |
| 2020-01-11 | 119.64 ± 8.27 | $r'$ | 2020-02-15 | 126.68 ± 0.90 | $r'$ |

The results presented in Figure 8 and Table 6 demonstrate a significant positive correlation between the color index and magnitude, with the exception of specific dates: 2018 December 26; 2019 January 8, 26; 2019 February 2; 2019 February 4, 14; 2018 December 19, 24, 25, 26; 2020 January 7, 9, and 25; and 2020 February 3, 14, 15. This suggests that the source has a BWB color trend on most nights, which is consistent with those results reported by some authors who found that the source exhibits a strong BWB trend (e.g., J. Wu et al. 2007; H. Poon et al. 2009; J. Wu et al. 2012; S. M. Hu et al. 2014; A. Agarwal et al. 2015; G. Bhatta et al. 2016a; S. Hong et al. 2017; C.-J. Wang et al. 2019; H.-C. Feng et al. 2020a, 2020b; Y. Dai et al. 2021). Furthermore, there is no significant correlation observed between color and magnitude during certain periods, thereby suggesting the absence of any trend in color change or spectral variability exhibited by the source within these time intervals. Such achromatic color trends may be related to a variation of the Doppler factor (M. Villata et al. 2004; M.-F. Gu & Y. L. Ai 2011; S. M. Hu et al. 2014; D. Xiong et al. 2020). The emission flux density $F_\nu$ and the frequency $\nu$ change as the Doppler factor changes. If the intrinsic source spectrum follows a power law, variations in the Doppler factor will not lead to observable changes in color trends. Moreover, the complex color behavior of blazars is influenced by various factors including superposition of multiple new variable components (H. Gaur et al. 2015), electron energy distribution (A. Mastichiadis & J. G. Kirk 2002), ratio between accretion disk and jet contributions (J. C. Isler et al. 2017), time lag (J. Wu et al. 2007), and Doppler factor variations (M. Villata et al. 2002, 2004; I. E. Papadakis et al. 2007).

The behavior of the BWB can be attributed to nonthermal synchrotron emission from the relativistic jet and explained under the scenario of the shock-in-jet model (e.g., A. P. Marscher & W. K. Gear 1985; G. Ghisellini et al. 1997; J. G. Kirk et al. 1998; A. Mastichiadis & J. G. Kirk 2002; A. P. Marscher et al. 2008; H. Poon et al. 2009; M.-F. Gu & Y. L. Ai 2011; Y. Ikejiri et al. 2011; B.-z. Dai et al. 2015; A. Wierzcholska

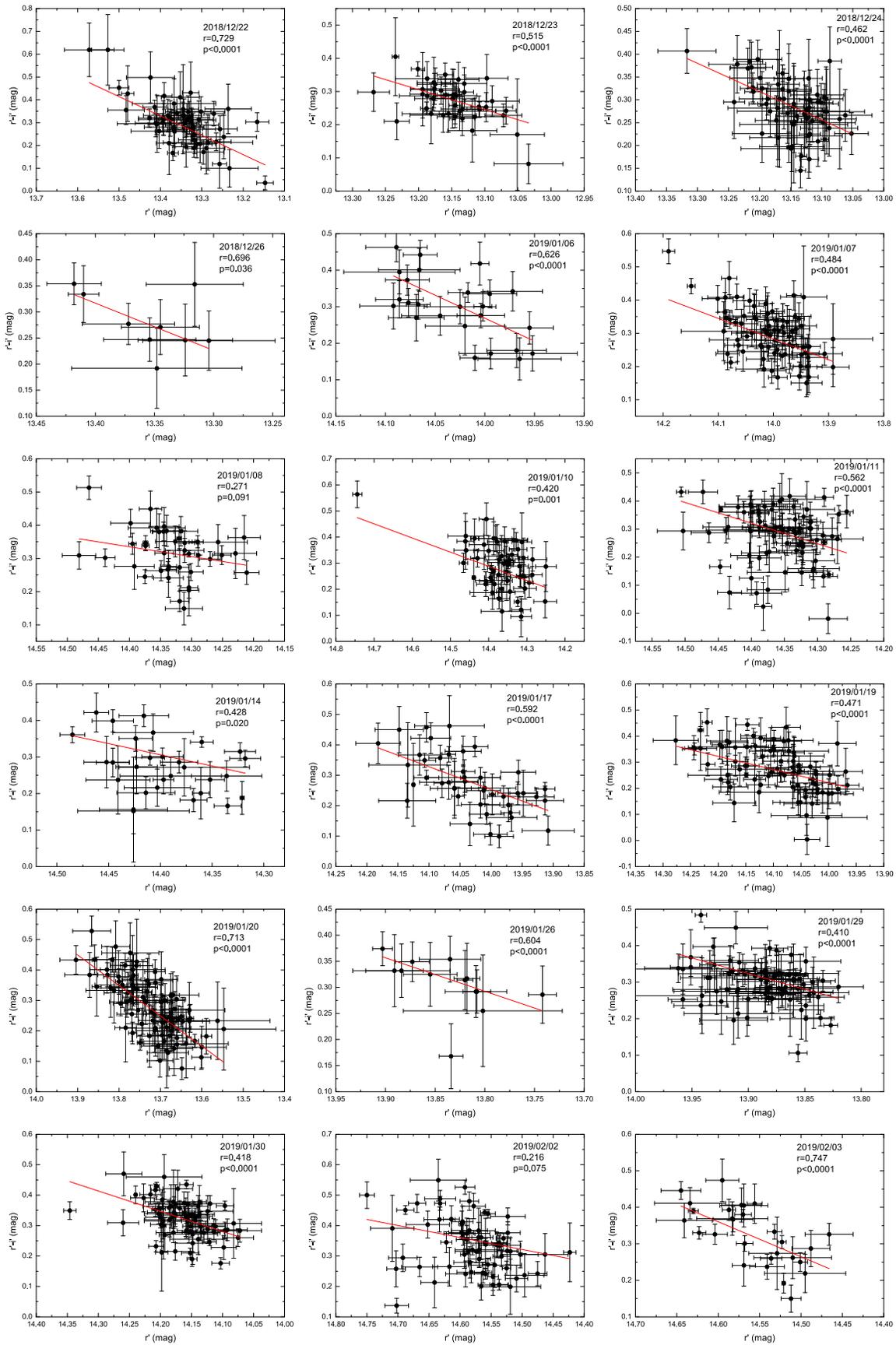

**Figure 8.** The color–magnitude diagrams for each separate night observation of S5 0716+714. The red line is the linear fitting.

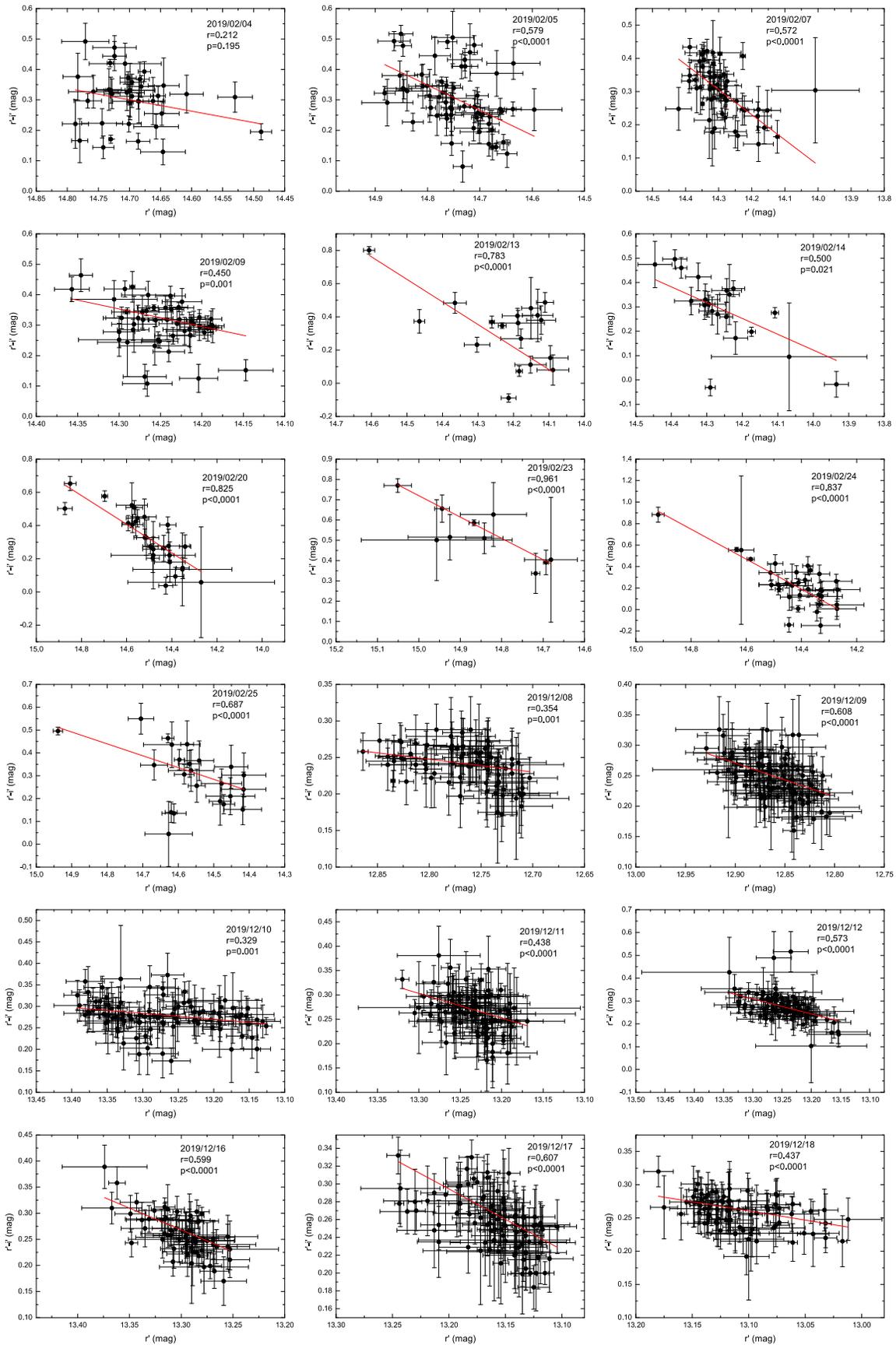

**Figure 8.** (Continued.)

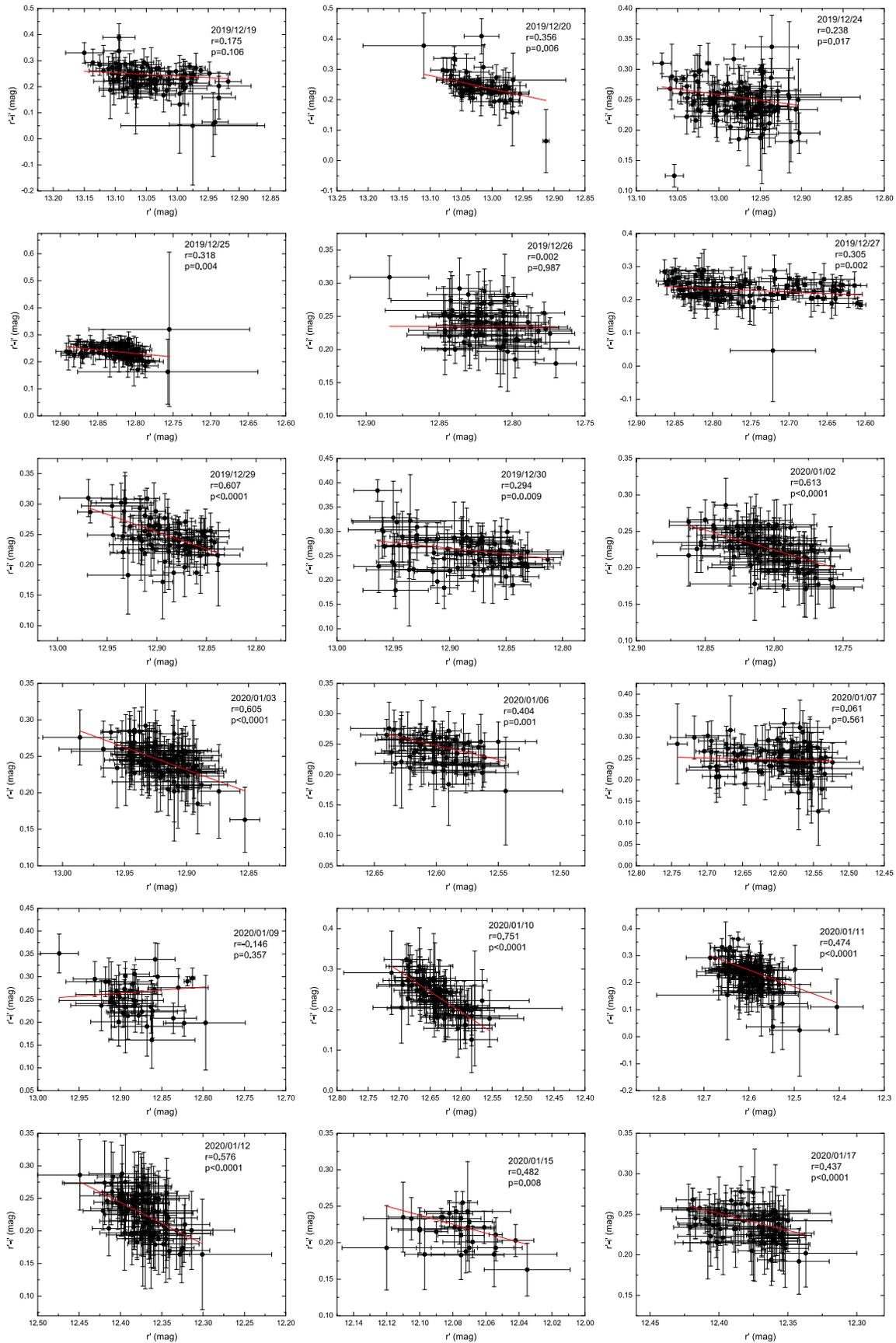

**Figure 8.** (Continued.)

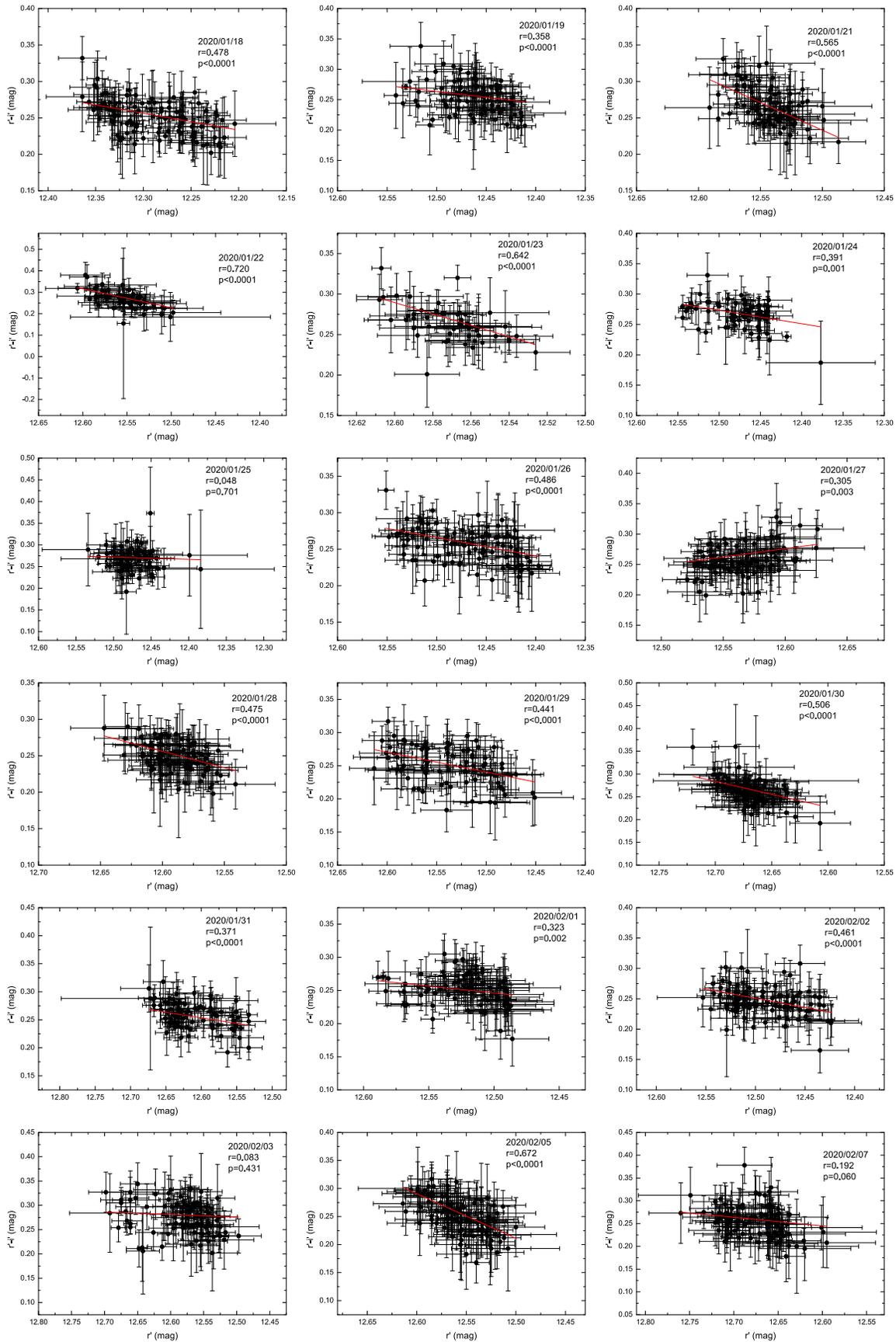

**Figure 8.** (Continued.)

THE ASTROPHYSICAL JOURNAL SUPPLEMENT SERIES, 277:18 (30pp), 2025 March                                                                    Li et al.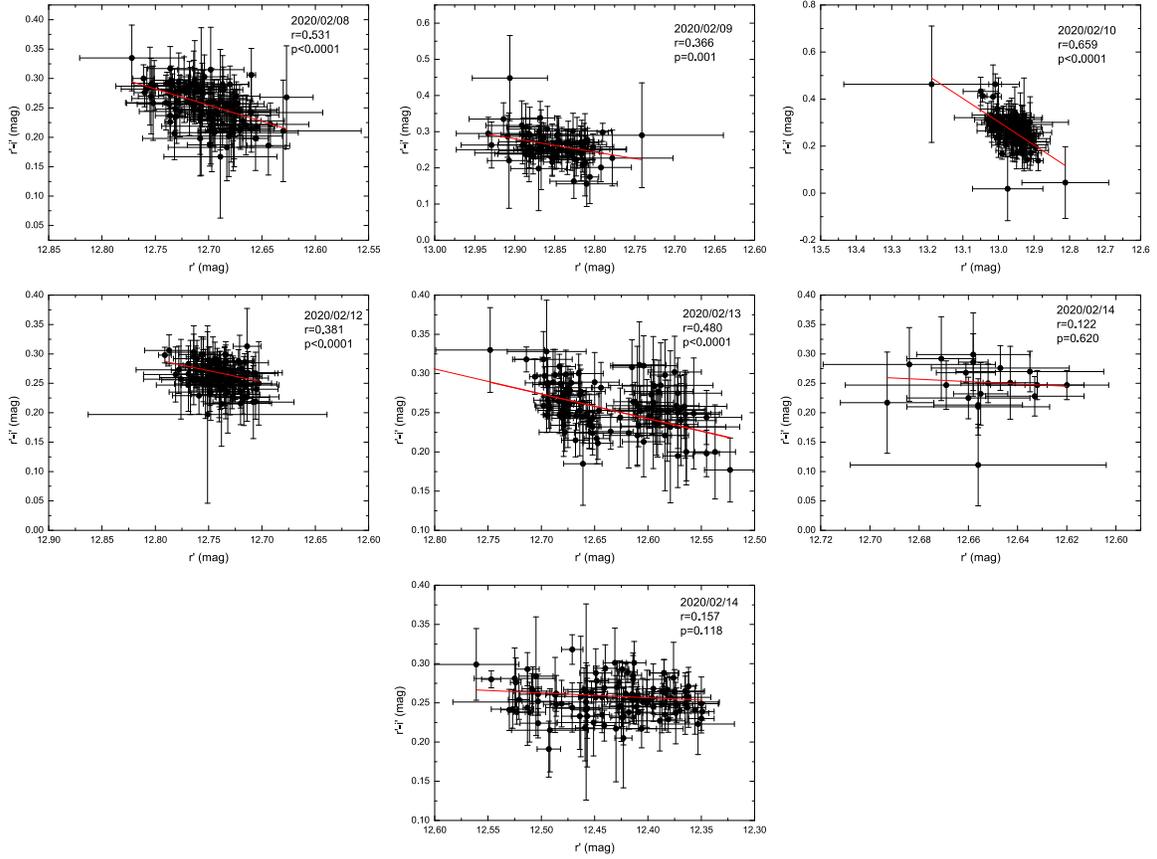

**Figure 8.** (Continued.)

et al. 2015; D. Xiong et al. 2016, 2017; S. Hong et al. 2017; H.-C. Feng et al. 2020a; V. Negi et al. 2022; H.-Z. Li et al. 2024a; T. Tripathi et al. 2024). In the shock-in-jet model, a shock is generated at the base of the relativistic jet and propagates along it. The shock wave accelerates electrons and compresses the magnetic field at the front of the shock, and these processes will generate synchrotron radiation. Different wavelength radiations are produced at varying distances behind the shock. Higher frequency synchrotron radiation typically emerges more quickly and closer to the shock front. Consequently, when the blazar brightens, the higher-energy photons are emitted first or more intensely, leading to the BWB trend (A. Agarwal et al. 2015; S. Hong et al. 2017; D. Xiong et al. 2017; X. Chang et al. 2023). Moreover, the BWB trend could be associated with radiation cooling. In the shock-in-jet model, higher energy electrons produced at the front of a shock lose energy faster through radiation cooling. More efficient cooling by higher frequency radiation makes blazars become more variable at higher frequencies (J. G. Kirk et al. 1998; A. Mastichiadis & J. G. Kirk 2002; H. Poon et al. 2009). As a result, when the blazar brightens, the color becomes bluer because the higher energy electrons, which contribute more to the bluer photons, are cooling and changing the spectral energy distribution (V. Negi et al. 2022). The acceleration and subsequent cooling of electrons lead to a shift in $\nu_{\rm syn}^p$, which is influenced by both the relativistic electron distribution and the magnetic field (M.-F. Gu & Y. L. Ai 2011; Y. Ikejiri et al. 2011; H.-C. Feng et al. 2020a, 2020b). The position of $\nu_{\rm syn}^p$ relative to the observational frequency ranges will determine the detected spectral index (H.-C. Feng et al. 2020a, 2020b). When blazars brighten, the shock acceleration efficiency should surpass the radiation cooling, resulting in an increase in $\nu_{\rm syn}^p$. If the observational frequency ranges fall to the right of $\nu_{\rm syn}^p$, the BWB behavior will be observed for S5 0716+714 (H.-C. Feng et al. 2020b).

### 4.4. Long-term Optical Spectral Variability

The optical emission is typically generated through synchrotron radiation processes and can be described by a typical synchrotron power law, namely, $F_\nu \propto \nu^{-\alpha}$, where $F_\nu$ and $\alpha$ represent the flux and the spectral index at frequency $\nu$. Therefore, the optical spectral index $\alpha$ can be obtained through a linear least-squares fit of $\log F_\nu$ versus $\log \nu$ (Y. G. Zheng et al. 2008). To achieve a more optimal fit, we select quasi-simultaneous data comprising three or more bands. The retained results will only include those with a Pearson's linear correlation coefficient greater than 0.9 and a standard deviation of the slope less than 0.5, ensuring the precision of the spectral index (Y. G. Zheng et al. 2008). The logarithm of V-band flux $\log F_V$ and the spectral index are plotted in Figure 9.

#### 4.4.1. Correlation between Flux and Spectral Index

The intraday color behavior suggests that S5 0716+714 exhibits BWB behavior, which has also been reported by many authors (e.g., H. Poon et al. 2009; G. Bhatta et al. 2016a; S. Hong et al. 2017; C.-J. Wang et al. 2019; H.-C. Feng et al. 2020a). In order to further study spectral variability, the correlation between the spectral index and V-band flux density has been analyzed. The correlation between $\log F_V$ and $\alpha$ is shown in Figure 10. The correlation was obtained through the utilization of linear regression, and the regression equation is

Table 6
Results of Error-weighted Linear Regression Analysis

| Date (1) | N (2) | r (3) | p (4) | Date (1) | N (2) | r (3) | p (4) |
|---|---|---|---|---|---|---|---|
| 2018-12-22 | 59 | 0.729 | <0.0001 | 2019-12-26 | 69 | 0.002 | 0.987 |
| 2018-12-23 | 42 | 0.515 | <0.0001 | 2019-12-27 | 101 | 0.305 | 0.002 |
| 2018-12-24 | 62 | 0.462 | <0.0001 | 2019-12-29 | 77 | 0.607 | <0.0001 |
| 2018-12-26 | 9 | 0.698 | 0.036 | 2019-12-30 | 79 | 0.294 | 0.009 |
| 2019-01-06 | 25 | 0.626 | <0.0001 | 2020-01-02 | 96 | 0.613 | <0.0001 |
| 2019-01-07 | 74 | 0.484 | <0.0001 | 2020-01-03 | 97 | 0.605 | <0.0001 |
| 2019-01-08 | 40 | 0.271 | 0.091 | 2020-01-06 | 61 | 0.404 | 0.001 |
| 2019-01-10 | 56 | 0.420 | 0.001 | 2020-01-07 | 92 | 0.061 | 0.561 |
| 2019-01-11 | 77 | 0.562 | <0.0001 | 2020-01-09 | 42 | −0.146 | 0.357 |
| 2019-01-14 | 29 | 0.428 | 0.020 | 2020-01-10 | 74 | 0.751 | <0.0001 |
| 2019-01-17 | 43 | 0.592 | <0.0001 | 2020-01-11 | 105 | 0.474 | <0.0001 |
| 2019-01-19 | 79 | 0.471 | <0.0001 | 2020-01-12 | 95 | 0.576 | <0.0001 |
| 2019-01-20 | 91 | 0.713 | <0.0001 | 2020-01-15 | 29 | 0.482 | 0.008 |
| 2019-01-26 | 13 | 0.604 | 0.029 | 2020-01-17 | 78 | 0.437 | <0.0001 |
| 2019-01-29 | 77 | 0.410 | <0.0001 | 2020-01-18 | 105 | 0.478 | <0.0001 |
| 2019-01-30 | 75 | 0.418 | <0.0001 | 2020-01-19 | 86 | 0.358 | <0.0001 |
| 2019-02-02 | 69 | 0.216 | 0.075 | 2020-01-21 | 82 | 0.565 | <0.0001 |
| 2019-02-03 | 28 | 0.747 | <0.0001 | 2020-01-22 | 56 | 0.720 | <0.0001 |
| 2019-02-04 | 39 | 0.212 | 0.195 | 2020-01-23 | 50 | 0.642 | <0.0001 |
| 2019-02-05 | 62 | 0.579 | <0.0001 | 2020-01-24 | 66 | 0.391 | 0.001 |
| 2019-02-07 | 53 | 0.572 | <0.0001 | 2020-01-25 | 67 | 0.048 | 0.701 |
| 2019-02-09 | 50 | 0.450 | 0.001 | 2020-01-26 | 97 | 0.486 | <0.0001 |
| 2019-02-13 | 18 | 0.783 | <0.0001 | 2020-01-27 | 96 | 0.305 | 0.003 |
| 2019-02-14 | 21 | 0.500 | 0.021 | 2020-01-28 | 95 | 0.475 | <0.0001 |
| 2019-02-20 | 27 | 0.825 | <0.0001 | 2020-01-29 | 82 | 0.441 | <0.0001 |
| 2019-02-23 | 10 | 0.961 | <0.0001 | 2020-01-30 | 89 | 0.506 | <0.0001 |
| 2019-02-24 | 33 | 0.837 | <0.0001 | 2020-01-31 | 80 | 0.371 | <0.0001 |
| 2019-02-25 | 23 | 0.687 | <0.0001 | 2020-02-01 | 90 | 0.323 | 0.002 |
| 2019-12-08 | 78 | 0.354 | 0.001 | 2020-02-02 | 94 | 0.461 | <0.0001 |
| 2019-12-09 | 96 | 0.608 | <0.0001 | 2020-02-03 | 93 | 0.083 | 0.431 |
| 2019-12-10 | 95 | 0.329 | 0.001 | 2020-02-05 | 99 | 0.672 | <0.0001 |
| 2019-12-11 | 91 | 0.438 | <0.0001 | 2020-02-07 | 97 | 0.192 | 0.060 |
| 2019-12-12 | 79 | 0.573 | <0.0001 | 2020-02-08 | 104 | 0.531 | <0.0001 |
| 2019-12-16 | 62 | 0.599 | <0.0001 | 2020-02-09 | 73 | 0.366 | 0.001 |
| 2019-12-17 | 90 | 0.607 | <0.0001 | 2020-02-10 | 101 | 0.659 | <0.0001 |
| 2019-12-18 | 76 | 0.437 | <0.0001 | 2020-02-12 | 94 | 0.381 | <0.0001 |
| 2019-12-19 | 87 | 0.175 | 0.106 | 2020-02-13 | 106 | 0.480 | <0.0001 |
| 2019-12-20 | 58 | 0.356 | 0.006 | 2020-02-14 | 19 | 0.122 | 0.620 |
| 2019-12-24 | 101 | 0.238 | 0.017 | 2020-02-15 | 101 | 0.157 | 0.118 |
| 2019-12-25 | 81 | 0.318 | 0.004 | | | | |

given by the following equation:

$$\alpha = (-0.601 \pm 0.012)\log F_V + (1.798 \pm 0.015), \quad (16)$$

with a Pearson coefficient of $r = 0.737$, and a chance probability of $p < 10^{-4}$. This suggests that there is a strong anticorrelation between the spectral index and the V-band flux density. The evidence from Figure 10 further substantiates the BWB behavior exists on S5 0716+714.

The phenomenon of spectral variation is widely observed in BL Lacertae objects, and it can arise from various underlying factors (e.g., M. Villata et al. 2007; Y. G. Zheng et al. 2008; H. Poon et al. 2009; S. M. Hu et al. 2014; J. C. Isler et al. 2017; D. Xiong et al. 2020). For example, the spectral behavior may be related to the particle acceleration and cooling mechanisms. If the electrons are accelerated to preferentially higher energies before radiative cooling, the spectra will exhibit a flatter-when-brighter trend. Moreover, if the highest-energy electrons suffer a stronger radiative cooling or escape cooling, the spectral behavior will exhibit steeper-when-fainter trends (J. C. Isler et al. 2017; D. Xiong et al. 2020). Moreover, according to M. Fiorucci et al. (2004), it was observed that the energy of particles increases with a higher intensity of energy release, resulting in a tendency for brighter emissions to be flatter. Furthermore, if the increase in luminosity is attributed to fresh electron injection, the resulting energy distribution exhibits a harder spectrum compared to the previously partially cooled state (J. G. Kirk et al. 1998; A. Mastichiadis & J. G. Kirk 2002; Y. G. Zheng et al. 2008).

*4.4.2. Periodicity of Flux and Spectral Index Variability*

The periodicity variability with a variety of timescales of the emission of S5 0716+714 has been reported by many authors (e.g., C. M. Raiteri et al. 2003; N. H. Liao et al. 2014; Y.-H. Yuan et al. 2017; H. Z. Li et al. 2018; X.-P. Li et al. 2023; H. Yang et al. 2023). Figure 9 shows that the optical spectral index variability exhibits a similar pattern to that of the V-band flux density, indicating a potential correlation between them. Consequently, there might be periodicity in the variation of the optical spectral index. In order to explore the periodicity of the emission variability and the spectral index, the REDFIT program was employed to analyze the variability data of the optical *V* band and spectral index. The REDFIT program enables direct estimation of red noise spectra from unevenly spaced time series (M. Schulz & M. Mudelsee 2002). The REDFIT program was developed based on the first-order autoregressive (AR1) model, which is included in the generally used and more robust ARIMA (p, d, q) test with AR (p) and MA(q), d = 0,1, etc. It is often performed to estimate the red noise spectrum from the data time series by fitting a first-order autoregressive process. The program is capable of directly analyzing unevenly spaced time series, thereby avoiding interpolation in the time domain and its inevitable bias. If a peak in the spectrum of a time series is significant against the red noise background resulting from an AR1 process, it indicates the presence of a significant variability timescale. Moreover, the REDFIT program can calculate the significance of the result, and provide the false alarm probability levels of the result with a maximum $2.5\sigma$ (99%).

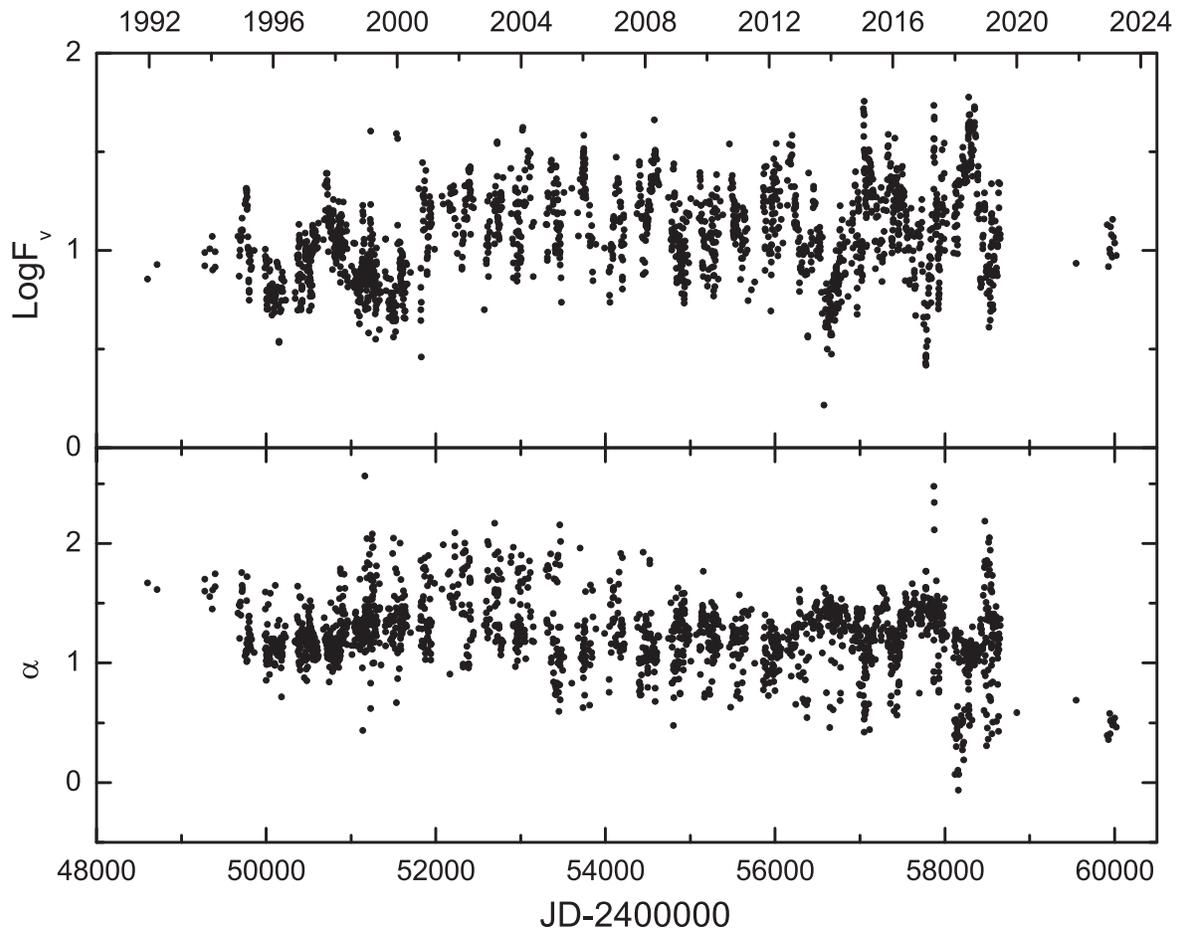

**Figure 9.** The *V*-band light curve and optical spectral curve of S5 0716+714.

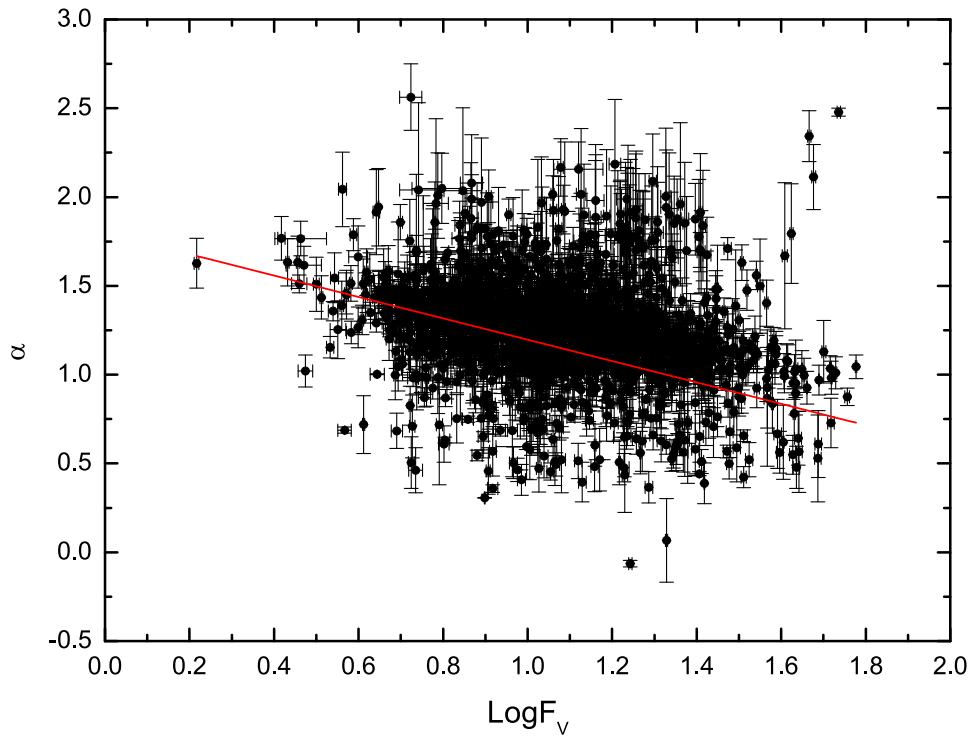

**Figure 10.** Long-term optical spectral index vs. *V*-band flux for S5 0716+714. The red line represents the regression line, characterized by a slope of $-0.601 \pm 0.012$, an intercept of $1.798 \pm 0.015$, and a Pearson coefficient of $-0.737$.

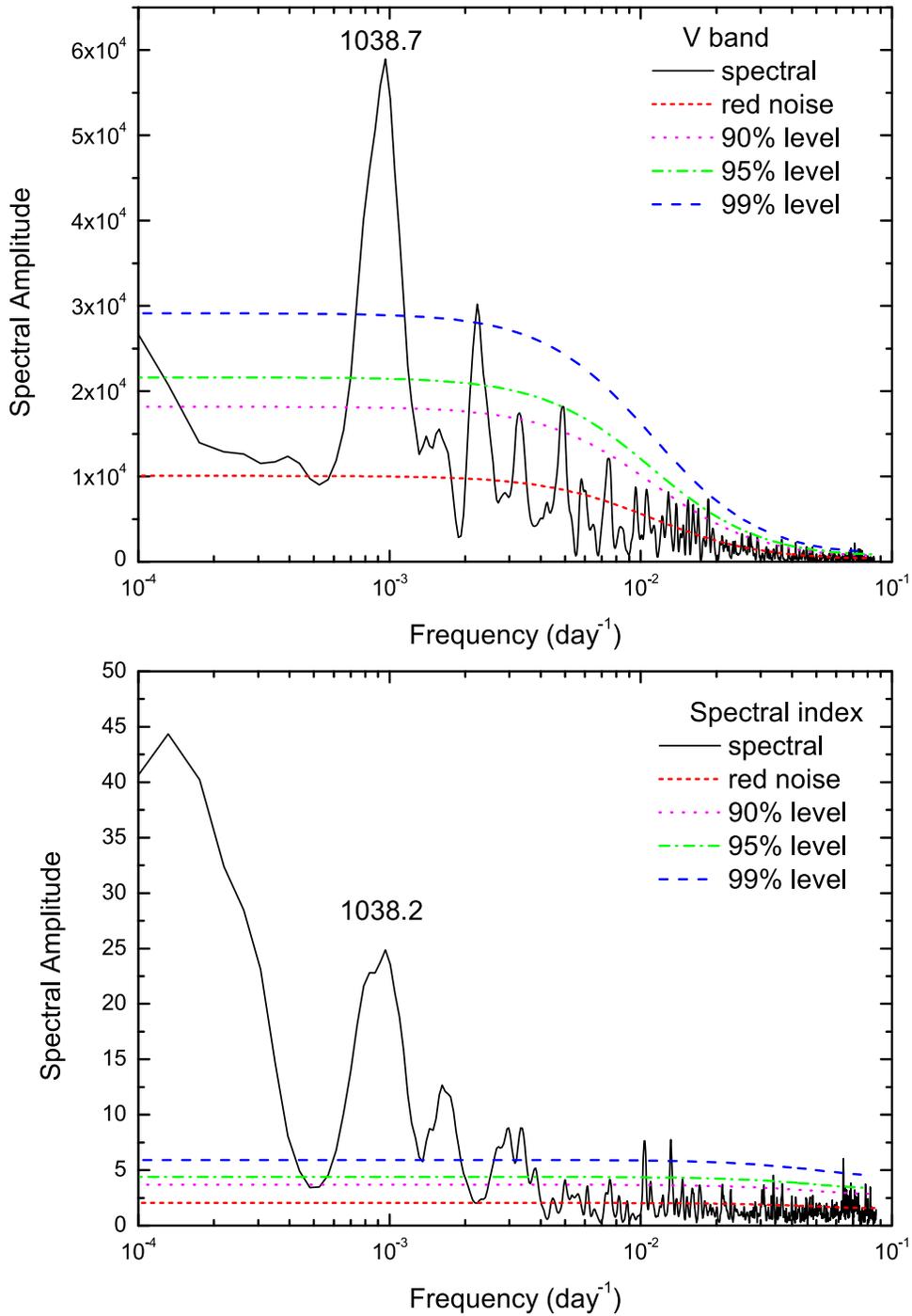

**Figure 11.** The REDFIT program analysis results for the *V*-band flux and optical spectral index of S5 0716+714. The top and bottom panels represent the results of the optical *V* band and spectral index, respectively.

The REDFIT program is used to analyze the variability timescale of the *V*-band flux density and optical spectral index. The results are plotted in Figure 11, where the top and bottom panels represent the REDFIT results of the optical *V* band and spectral index, respectively. The top panel of Figure 11 exhibits a significant peak at a timescale of 1038.7 days over the red noise spectrum of the V-band flux density, suggesting that S5 0716 +714 exhibits a significant quasiperiodicity in the optical band with a timescale of 1038.7 days. This period is completely consistent with the result reported by H. Yang et al. (2023), who found that there is a prominent quasiperiodic oscillation of 1060 days. Moreover, the bottom panel of Figure 11 also reveals a prominent quasiperiodicity in the optical spectral index variability, exhibiting a timescale of 1038.2 days, which is in good agreement with the variability period in the V-band flux density.

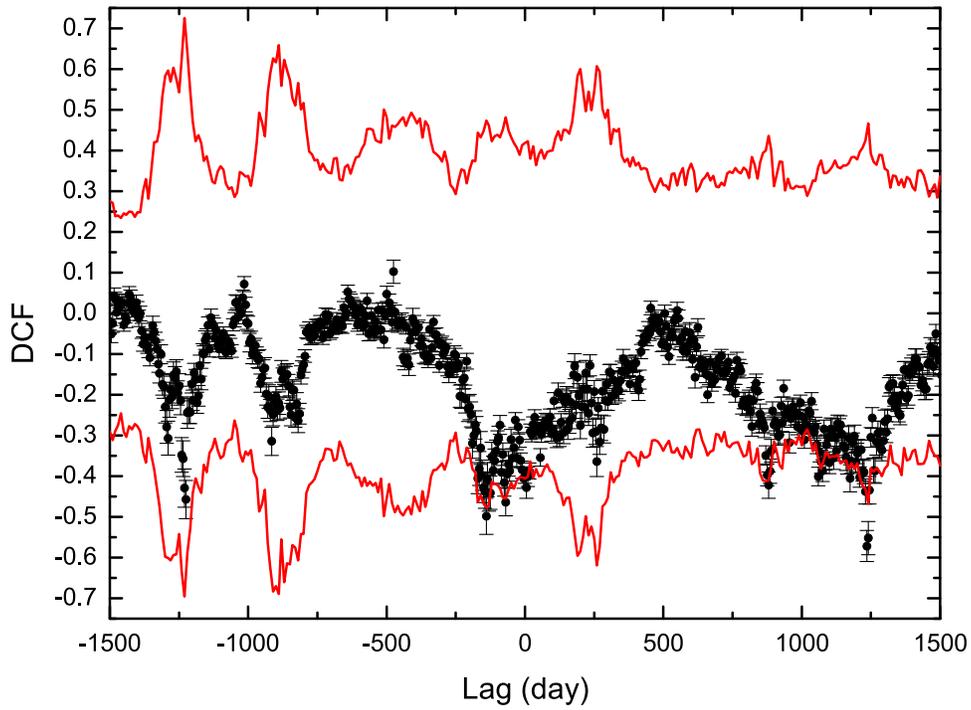

**Figure 12.** The time lag between the optical variability and optical spectral index with the DCF. The red dashed line represents a confidence level of $3\sigma$.

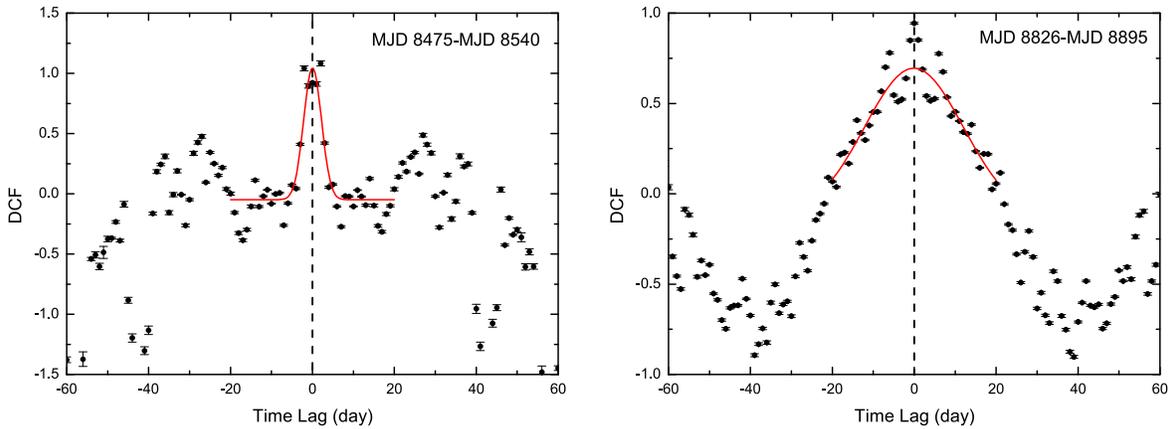

**Figure 13.** The result of correlation analysis between the Sloan $i'$- and $r'$-band light curves. The left and right picture is the result based on the observation data during MJD 8475-8540 and MJD 8826-8895, respectively. The red line is the Gaussian fitting to measure the time lag.

Since there are variability periodicities around 1038 days in both the optical flux and spectral index, it is natural to wonder if these two changes are synchronous. The discrete correlation function (DCF) between the optical flux and spectral index in Figure 12 shows a time lag of −140 days at a DCF value of −0.50 ± 0.04. This anticorrelation has a confidence level exceeding 99.7% (>$3\sigma$). Monte Carlo simulations, following the methodology established in previous studies (X. Yang et al. 2020; Y.-F. Wang & Y.-G. Jiang 2021), were run to give the confidence level curves for correlation coefficients in the DCF (see Figure 12). These two asynchronous periodic variations are an interesting phenomenon and might result from a structured jet with a periodically changing viewing angle, which might be from helical motion driven by the orbital motion in a binary black hole system and helical jet paths driven by jet precession (F. M. Rieger 2004). A structured jet with a fast spine surrounded by a slower layer was used to explain the broadband spectrum from optical to VHE emission for blazars (e.g., F. Tavecchio & G. Ghisellini 2008, 2014, 2015). The optical spectrum index should be controlled by the relative strengths and spectral shapes of the two radiation components from the spine and layer structures of the jet. When the Doppler factors of the spine and layer of jet change synchronously due to the wiggling jet, it is possible that the Doppler factors, intrinsic optical flux densities, and spectral indices do not change synchronously, e.g., asynchronously reaching maxima or minima. Also, it is possible that the wiggling jet without transverse velocity structure can give rise to these two asynchronous periodic variations. Though numerical simulation is needed to give interpretation to the physical origin of these two asynchronous periodic variations, it should be out of the scope of this work. After correcting the cosmological redshift effect, the time lag of 140 days is 107 days.

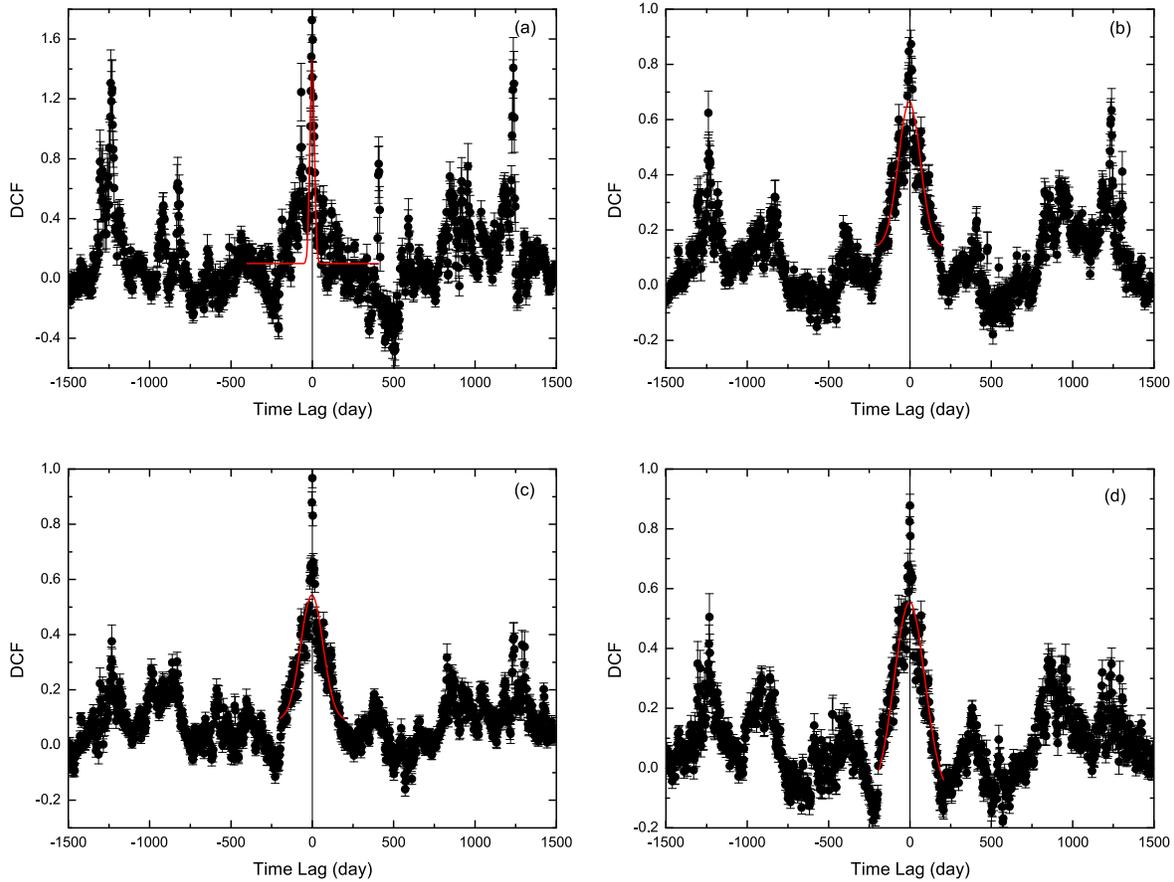

**Figure 14.** The result of long-term correlation analysis between the UBRI- and V-band light curves. The red line is the Gaussian fitting to measure the time lag. The correlation between the V and U bands, V and B bands, R and V bands, and I and V bands are depicted in (a), (b), (c), and (d), respectively.

### 4.5. Correlation Analysis and Time Lag

The variability correlation and time lag between the interbands can help us to understand the emission mechanisms (H.-Z. Li et al. 2022b). The significant correlation observed among the different bands suggests a consistent radiation mechanism between them. Moreover, the absence of correlation among the various bands of radiation suggests that there exists a distinct radiation mechanism for each band. Furthermore, the time lag is correlated with variations in the radiation region. Based on our observed data on the $i'$ and $r'$ bands, we investigate the correlation and time lag between the variability of the $i'$ and $r'$ bands using the DCF method. In addition, we also study the long-term correlation and time lag between the variabilities of different optical bands.

The algorithm of the DCF method is given by R. A. Edelson & J. H. Krolik (1988). From the DCF profile bumps closer to zero lag, the centroid time lags τ are used to estimate the time lag. For the observation data, where are the two distinct periods of observation (see Figure 1). Based on the observation period, we utilized the DCF method to examine the radiation correlation and time delay between the $i'$ and $r'$ bands in two distinct segments (as illustrated in Figure 2). Moreover, the Gaussian fitting was made to the central DCF results to confirm the timescale of the time lag. The results were plotted in Figure 13, indicating a significant correlation between the variability of the $i'$ and $r'$ bands without any discernible time delay. The long-term correlation between the variability of the UBRI and V bands is further analyzed using the DCF method based on the optical variability data plotted in Figure 3. The results, presented in Figure 14, exhibit a prominent peak at zero, indicating a significant long-term correlation between the UBRI and V bands with no time lag. These findings corroborate previous studies by various authors (e.g., S. Hong et al. 2017) and suggest the congruence of emission progress across different optical bands. The frequency stratification observed in shock-in-jet models offers an explanation for the time delay between different wave bands. In this model, a relativistic shock propagates within the jet, causing the emission region to move along the jet at a bulk speed of $\beta c$ (W. Max-Moerbeck et al. 2014; H. Z. Li et al. 2016). During this motion, regions closer to the wave front exhibit higher frequency emissions. Consequently, high-frequency radiation is detected prior to low-frequency radiation. No discernible time delay is observed among the UBVRI optical bands, as well as between the $i'$ and $r'$ bands, due to their narrow coverage within the electromagnetic spectrum.

## 5. Conclusion

We have monitored S5 0716+714 in the Sloan $i'$ and $r'$ bands from 2018 December 22 to 2020 February 15, obtaining more than 5600 effective images on each filter over a span of 79 nights. Moreover, we have compiled long-term multiband variability data spanning over 34 yr and subsequently computed the optical spectral index. The main summary of our results is as follows:

(1) During the 79 nights, IDV was detected on 31 nights and possible IDV on 20 nights in the $i'$ band. Similarly, IDV was detected on 35 nights in the $r'$ band, while possible IDV was observed on 22 nights.

(2) The minimum timescale of variability was estimated using ACF, ranging from $7.33 \pm 0.74$ minutes. Additionally, we determined the emission region within the jet and calculated black hole masses, which spanned a range of $0.4 \times 10^{14}$ to $3.02 \times 10^{14}$ cm and $0.68 \times 10^{8}$ to $5.12 \times 10^{8} M_{\odot}$, respectively.

(3) The spectral variability was investigated, revealing that S5 0716+714 exhibited BWB color behavior during our observation period. Moreover, the optical flux density and spectral index displayed long-term variations with a periodicity of approximately 1038 days, while a significant anticorrelation between the spectral index and optical flux density was observed. Additionally, there existed a time delay of $-140$ days between the variations in optical flux density and spectral index.

(4) The analysis of the correlation and time delay between our observed data and long-term optical variability revealed a strong correlation across various optical bands, with no significant time lag.

### Acknowledgments

We owe great thanks to the Xingming Observatory staff, who contributed great efforts to this campaign. This work is supported by the National Natural Science Foundation of China (12373018, 12063005, 12063006), The Young and Middle-aged Discipline and Technology Leaders Reserve Talents in Yunnan Province (202405AC350114). The authors gratefully acknowledge the computing support provided by the JRT Science Data Center at Yuxi Normal University and the authors (Q.L.H.) gratefully acknowledge the financial support from the Hundred Talents Program of Yuxi (grants 2019).

### ORCID iDs

Huai-Zhen Li https://orcid.org/0000-0001-8307-1442
Di-Fu Guo https://orcid.org/0000-0003-4957-485X
Long-Hua Qin https://orcid.org/0000-0001-7905-4295
Hong-Tao Liu https://orcid.org/0000-0002-2153-3688
Ting-Feng Yi https://orcid.org/0000-0001-8920-0073
Quan-Gui Gao https://orcid.org/0000-0001-9732-069X
Shi-Feng Huang https://orcid.org/0000-0001-7689-6382
Xu Chen https://orcid.org/0000-0001-5603-7521